\documentclass[reprint,prl,notitlepage,
superscriptaddress,aps]{revtex4-1}
\usepackage[utf8]{inputenc}
\usepackage{braket}
\usepackage{amsmath}
\usepackage{amsthm}
\usepackage{mathrsfs}
\usepackage{mathtools}
\usepackage{amssymb}
\usepackage{dsfont}
\usepackage{bm}
\DeclareMathOperator{\tr}{tr}
\newtheorem{thm}{Theorem}[]

\newtheorem{lemma}[thm]{Lemma}

\usepackage{soul,xcolor}
\begin{document}

\title{Detecting Quantum Capacities of Continuous-Variable Quantum Channels}

\author{Ya-Dong Wu}
\affiliation{QICI Quantum Information and Computation Initiative, Department of Computer Science,
The University of Hong Kong, Pokfulam Road, Hong Kong}
\author{Giulio Chiribella}
\email{giulio@cs.hku.hk}
\affiliation{QICI Quantum Information and Computation Initiative, Department of Computer Science,
The University of Hong Kong, Pokfulam Road, Hong Kong}
\affiliation{Department of Computer Science, Parks Road, Oxford, OX1 3QD, United Kingdom}
\affiliation{Perimeter Institute for Theoretical Physics, Waterloo, Ontario N2L 2Y5, Canada}

\begin{abstract}
Quantum communication channels and quantum memories  are the  fundamental building blocks of large-scale quantum communication  networks. Estimating their  capacity  to transmit and store quantum information is crucial in order  to assess the performance of quantum communication systems, and to detect useful communication paths among the nodes of future quantum networks. 
However, the estimation of quantum capacities is a  challenging task for continuous variable systems, such as the radiation field,  for which a complete  characterization  via quantum tomography is practically unfeasible.   Here we introduce  a method for detecting the quantum  capacity of continuous variable communication channels and memories without performing a full process tomography.  Our method works in the general scenario where the devices  are used a finite number of times, can exhibit correlations across multiple uses, and can change dynamically  under the control of a malicious adversary. The method  is  experimentally friendly and can be  implemented using only finitely-squeezed states  and homodyne measurements.  
\end{abstract}

\maketitle

{\em Introduction.}
 Continuous variable (CV) quantum systems are  a promising platform for the realization of quantum technologies, including quantum communication~\cite{Braunstein1998,furusawa1998,jouguet2013,pirandola2015,pirandola2017}, quantum computation~\cite{Nico06,Mile09,Ben19}, and the quantum internet~\cite{o2009photonic,RevModPhys.84.621}. 
 An essential building block for all these quantum technologies is the realization of devices that reliably transmit or store quantum information~\cite{gottesman2001,mirrahimi2014,marios16,victor2018,noh2018,sharma2018,noh2020,noh2020(1)}.
 An important performance measure for these devices is the quantum capacity~\cite{lloyd97,shor2002,devetak2005,nielsen2010,wilde2013}, that is, the number of  qubits that can be transmitted or stored with each use of the device under consideration.  To assess the performance of realistic devices, one needs methods to estimate the  capacity from experimental data.  Such methods are important not only for the certification of new quantum hardware, but also as a way to monitor future  quantum communication networks, in which the quality and availability of communication links may change dynamically due to fluctuations in the environment or to the amount of network traffic.   In this setting, the estimation of the quantum capacity provides  a way to assess how much information can be transmitted from a node to another during a given time frame, and to  identify optimal paths  for routing quantum information through the network. 

%one can hope to transmit or store by setting up suitable encoding and decoding operations. 
% From a fundamental point of view, estimating quantum capacity between any input and any output is a rigorous way to identify a quantum signalling pattern in a large-scaled quantum network, and hence provides an important approach to identify quantum causal strengths among quantum systems~\cite{costa2016,giarmatzi2018,bai2020}.
 
Unfortunately,  explicit expressions for the quantum capacity are  only known for particularly simple noise models, under the assumption that the noise processes at different times are independent and identically distributed  (i.i.d.)~\cite{giovannetti05,wolf071,Holevo01,wolf07}.  In realistic scenarios,  however, the noise can change over time and can exhibit correlations across different uses of the same device~\cite{RevModPhys.86.1203}.  Moreover, the calculation of the quantum capacity requires a classical description of the devices under consideration. To obtain such a description, one generally needs a full quantum process tomography~\cite{chuang1997,Poyatos1997,dariano2001,dariano2003,Altepeter2003}, which however becomes practically unfeasible  for devices acting on high-dimensional quantum systems.

A promising approach to circumvent the above difficulties is to search for lower bounds on the quantum capacity, and for experimental setups that estimate such lower bounds without requiring a full process tomography. In this way, one can detect  a guaranteed   amount of quantum information that can be transmitted or stored.  For finite dimensional systems,  this approach has been explored in Refs.~\cite{macchiavello2016,macchiavello2016(1),cuevas2017}, which provided accessible lower bounds on the asymptotic quantum capacity  under the i.i.d. assumption.    For qubit channels, these results were extended in  Ref.~\cite{pfister2018} to a broader scenario involving a finite number of uses of the device, possibly exhibiting  correlations among different uses. However, the existing results do not apply to CV  quantum channels, due to the infinite dimensionality of input and output systems.

In this paper we introduce two protocols for the detection of quantum capacities in the CV domain.  The two protocols provide  experimentally accessible lower bounds on the number of qubits that can be transmitted or stored with a finite number of uses of a given CV device.  The first protocol works  in  the general  scenario  where the behaviour of the  device can change dynamically from one use to the next,  can  be under the control of a malicious adversary, and can exhibit correlations across different uses.  The second protocol works in the less challenging setting where the different uses of the  device are independent and identical. %It has a simpler experimental implementation and   a lower sample complexity, meaning that a smaller number of repetitions is sufficient to achieve a reliable estimate.
The protocol works for all phase-insensitive Gaussian channels~\cite{sharma2018} and requires only the preparation of coherent states. 
Both protocols can be implemented using current optical quantum technologies and provide a  practically useful method to validate quantum communication channels and quantum memories.

\begin{figure}
\begin{center}
\includegraphics[width=0.48\textwidth]{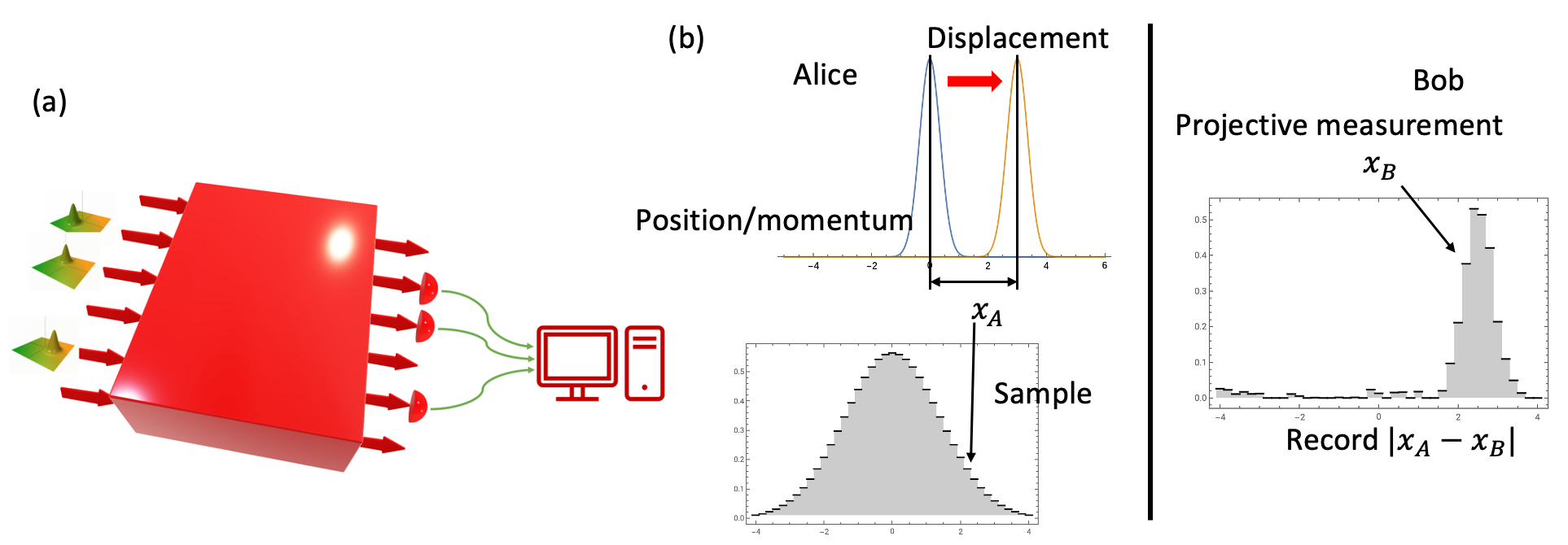}
\caption{
   (a) Capacity detection for continuous variable quantum channels. 
   The protocol deals with a completely unknown multimode quantum channel. A subset of the modes are randomly selected  for testing the channel, while the remaining modes are kept for communication. For each testing mode, the sender prepares a single-mode Gaussian input state. At the corresponding output port, a receiver performs a  Gaussian quantum measurement and sends the classical outcome to a classical computer for data analysis.  If the test is passed, then the sender and receiver  infer a lower bound on the quantum capacity of the channel acting on the communication modes.  For each communication mode, the sender can feed one part of a two-mode squeezed state into the  device,  keeping the other part for a later quantum communication task.
(b) Schematic diagram for Alice's and Bob's operations at each test mode in the first protocol.
%Here we choose $\eta=0.9$ and $V=5$ for the purple curve, while $\eta=0.75$ and $V=1$ for the orange curve.
}
\label{fig:schematic}
\end{center}
\end{figure}

Our protocols employ $k+n$ uses of the given quantum  device, and  randomly select $k$ uses for a test,  as shown in Fig.~\ref{fig:schematic}(a).  The test involves the preparation of single-mode input states (finitely-squeezed states in the first protocol, coherent states in the second) and the execution of single-mode  measurements on the output  (homodyne measurements in the first protocol, heterodyne in the second).  The result of the test is an estimated lower bound on the  number of qubits that can be transmitted with the remaining $n$ uses.    Notably, the sender and receiver do not need to agree in advance on which uses of the device will be employed for testing and which ones for communication: the sender can make this decision locally, and communicate it publicly after the transmission has taken place.

 In both protocols, the lower bound on the capacity comes hand in hand with a  lower bound on the amount of entanglement that can be established by  sending halves of two-mode finitely-squeezed vacuum states through the noisy channel under  consideration.
%Practically, the transmission can be achieved by feeding half of a two-mode squeezed state into each of the $n$  uses employed for communication, thus establishing entanglement between the sender and the receiver. 
By using the resulting entangled state as a resource, the sender and receiver can then achieve practical quantum communication, e.g. using optimal CV teleportation~\cite{pirandola2006,liuzzo2017}.
The quantum capacity is a lower bound on the private capacity, that is, the number of secret bits that can be sent reliably per channel use~\cite{horodecki2008}. For this reason, our estimated lower bound of the quantum capacity is also an estimated lower bound to the number of bits that can be sent privately through the channel. 
 
 {\em Background.  }
  A quantum process acting on a quantum system with Hilbert space $\cal H$   can be mathematically modeled by a quantum channel $\mathcal E: \mathcal S(\mathcal H)\rightarrow \mathcal S(\mathcal H)$, where ${\mathcal S} (\mathcal H)$ denotes the set of density operators on the Hilbert space  $\mathcal H$. 
%   For CV systems, the Hilbert space is infinite dimensional.  quantum harmonic oscillators, $\mathcal H=\operatorname{span}\{\ket{n}\}_{n=0}^\infty$ is an infinite-dimensional space. 
 The highest rate at which quantum information can be sent over a quantum channel $\cal E$ is quantified by its quantum capacity $Q(\mathcal E)$~\cite{chuang1997}.   The  definition of quantum capacity refers to the scenario where the channel  is used an asymptotically large number of times, and the noisy processes in the various uses of the channel  are identical and independently distributed.  In this scenario, the quantum capacity is defined as the maximum number of qubits that can be transmitted per use of the channel with optimal encoding and decoding maps, under the condition that the error must vanish in the asymptotic limit.

Practical applications, however, often deviate from the asymptotic  i.i.d scenario.  Noise can fluctuate in each run and correlations may arise between subsequent runs.   Realistically, the number of uses of the quantum channel is  always  finite,  and it is reasonable to allow for a finite error tolerance, as in the task of approximate quantum error correction~\cite{Leung1997,marios16,yang2020,philippe20,zhou2021}.   In these scenarios, it is convenient to adopt a one-shot version of the quantum capacity~\cite{buscemi2010}, denoted as  $Q^{\epsilon}(\mathcal E)$, where $\epsilon$ is the error tolerance. 
 Explicitly, the one-shot quantum capacity  is defined as
\begin{equation*}
Q^{\epsilon}(\mathcal E):=\max\{\log b | F(\mathcal E, b) \ge1-\epsilon\},
\end{equation*}
where $b$ is the dimension of the subspace in which information is encoded, and
\begin{equation*}
F(\mathcal E, b):=\max_{\bar{\mathcal H}\subset \mathcal H, \text{dim}(\bar{\mathcal H})=b} \max_{\mathcal D}\min_{\ket{\phi}\in\bar{\mathcal H}}\braket{\phi|\mathcal D\circ \mathcal E(\ket{\phi}\bra{\phi})|\phi},
\end{equation*}
is  the maximum  fidelity obtained by optimizing the choice of encoding subspace $\overline {\cal H}$ and the choice of a decoding channel $\cal D$, in the worst case over all possible input states. 
    When the channel is of the form $\mathcal E=\Lambda^{\otimes n}$, corresponding to $n$ i.i.d uses of a channel $\Lambda$, the asymptotic quantum capacity $Q(\Lambda)$ is equal to the limit of  the regularized one-shot  capacity $Q^{\epsilon}(\Lambda^{\otimes n})/n$   when  the number of uses goes to infinity and the error tolerance goes to zero.
In summary, the one-shot quantum capacity includes as a special case the asymptotic quantum capacity. 

In the following, we will consider the situation where  $\mathcal E$   acts on $n+k$ modes with Hilbert space  $\mathcal H^{\otimes (n+k)}$.     We will provide two protocols for experimentally estimating  lower bounds to the one-shot capacity. In the first protocol, the channel $\mathcal E$ will be an arbitrary $(n+k)$-mode channel, corresponding to the situation where the $n+k$ uses of the device are generally correlated. In the second protocol the channel will be assumed to be of the i.i.d form $\mathcal E   =  \Lambda^{\otimes (n+k)}$, where $\Lambda$ is a given single-mode channel, corresponding to the situation where the $n+k$ uses of the device are identical and independent.

{\em Protocol for arbitrary correlated noises.}
%The two methods presented in the previous section use two-mode entangled states and single-mode measurements. Equivalently, they  can be  formulated in terms of single-mode state preparations and measurements.
This protocol provides an experimentally accessible lower bound on the number of qubits that can be transmitted with a completely unknown multimode channel. The protocol can be viewed as an infinite-dimensional generalization of the approach of Ref.~\cite{pfister2018}. 
  A sender, Alice, prepares a quantum state of $k$ modes, each of which is subject to a finite amount of squeezing and displacement.  At the beginning, Alice randomly selects $k/2$ modes and initializes each of them in a single-mode position-squeezed vacuum state with finite amount of squeezing given by $s$ dB.    For the remaining  $k/2$ modes, she initializes them in  single-mode momentum-squeezed vacuum states with the same amount of squeezing. Practically, the amount of squeezing can be chosen by Alice depending on the experimental capabilities of her laboratory.   Then, Alice performs a random displacement on each mode, displacing the position-squeezed states (momentum-squeezed states) in position (momentum).  For each mode, the amount of displacement is chosen independently  according to a zero-mean Gaussian distribution with variance $\sigma^2  = 10^{\frac{s}{10}}$.   With this choice, the displaced squeezed states can be also  obtained by applying a homodyne measurement on one side of a two-mode squeezed state with finite mean photon number $\bar n=  (10^{\frac{s}{10}}-1)/2$.  

Notice that, while the variance is finite, there is still a non-zero probability that the randomly-chosen amount of displacement is too large to be implemented with Alice's devices.  To take this experimental limitation into account, we introduce a cut-off parameter $\alpha>0$ and allow Alice to repeat the randomization procedure until she gets a value in the interval $[-\alpha, \alpha]$.  The probability that Alice does not need  to repeat the randomization for a given mode is $p_{\alpha,s}  :=  \text{erf}\left( \alpha \,   10^{-\frac{s}{20}}/ \sqrt{2 }\right)$, where erf is error function.  Note that the probability $p_{\alpha, s}$ can be increased by increasing the amount of squeezing in the input states.

The receiver, Bob, performs homodyne detections on the $k$ modes sent by Alice. Specifically, Bob performs  position (momentum) measurements on the  $k/2$ position-displaced (momentum-displaced) modes.   Here we take into account  that in a realistic setting Bob's detectors will have a finite resolution, and therefore the measurement outcomes will be discretized. We denote by $d$ the width of the detector pixels in this discretization.

Finally, Alice and Bob perform a statistical test  of the correlations between Alice's displacements and Bob's outcomes.  For simplicity of analysis, we apply the cutoff $\alpha$ also to Bob's outcomes, and the discretization $d$ also to Alice's displacements. In this way, both outcomes and displacements become discrete dimensionless random variables in the finite interval $\{0, 1,\dots,  \frac{2\alpha}{d} -1\}$, having chosen   $2\alpha/d$ to be an integer.  In the following, we will denote by $\bm{x}_A$ ($\bm{x}_B$) the vector of Alice's displacements (Bob's outcomes).  The test is passed if the condition 
$\frac{1}{k}\sum_{i=1}^{k} |x_{A,i}- x_{B,i}|\le t$ is satisfied,
where $t$ is a threshold value chosen  by Alice and Bob. In the following, we will see that choosing smaller values of $t$ results into higher values of the capacity guaranteed by the test. On the other hand, however, low values of $t$ make the test harder to pass.

%Given any noisy quantum channel, the probability to pass the test can be boosted either by increasing the amount of squeezing at the input, or by raising the threshold $t$, which however reduces the lower bound on the capacity.

\begin{figure}
\begin{center}
\includegraphics[width=0.48\textwidth]{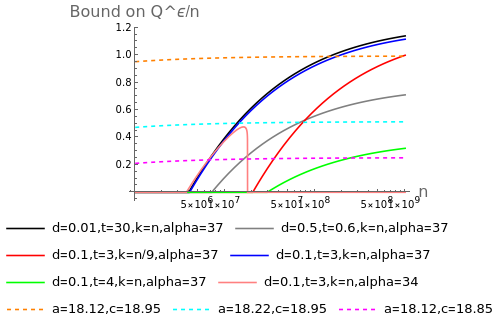}
\caption{ Solid curves are the lower bounds on $\frac{Q^{\epsilon}}{n}$, with $\epsilon=0.02$, given by Eq.~(\ref{eq:thm1}), as functions of~$n$ for different values of $d$, $t$, $k$ and $\alpha$, and dashed curves are the lower bounds on $\frac{Q^{\epsilon}}{n}$, given by Eq.~(\ref{eq:thm2}), as functions of~$n$ for different values of $a$ and $c$. 
Other parameters are $p_{\text{err}}=0.1$ and $\bar{n}=9.5$ for solid curves, and $k=n$ and $\bar{n}=9.5$ for dashed curves.
%Here we choose $\eta=0.9$ and $V=5$ for the purple curve, while $\eta=0.75$ and $V=1$ for the orange curve.
%Other parameters are  $\delta=0.01$, and $\bar{n}=9.5$ for dashed curves. 
%Here we choose $\eta=0.9$ and $V=5$ for the purple curve, while $\eta=0.75$ and $V=1$ for the orange curve.
}
\label{fig:scaling}
\end{center}
\end{figure}

\begin{thm}
If the test is passed  on $k$ randomly selected modes, then, with error probability no larger than $p_{\text{err}}$, 
 the  one-shot quantum capacity of the channel corresponding to other $n$ modes is lower bounded by  
\begin{equation}\label{eq:thm1}
Q^{\epsilon} \ge  \max\left\{ 0,    \sup_{\eta\in\left(0, \sqrt{\epsilon/2}-\lambda \right)}\,   f(\eta)  \right\},
\end{equation}
where \begin{align}
\nonumber &  f(\eta)  =   n \log  \frac{2\pi}{d^2}-2n\log  \gamma(t+\mu (\zeta (\eta)))- \Delta  (\eta)  \\
\nonumber    & \lambda=8\sqrt{2(1-  p_{\alpha,s}^n)}\left(3+\frac{5}{4p_{\text{err}}}-\frac{1}{\sqrt{p_{\text{err}}}}\right),\\
\nonumber &\gamma(x)=(x+\sqrt{1+x^2})\left(\frac{x}{\sqrt{1+x^2}-1}\right)^x,\\
\nonumber        & \mu (\zeta)=\frac{2\alpha}{d}\sqrt{\frac{(k+n)(k+1)}{nk^2}\log{\frac{1}{\zeta/4-2\sqrt{2(1-p_{\alpha,s}^n)}}}},\\
\nonumber & \zeta  (\eta)=\left(\sqrt{\frac \epsilon 2}-\eta+  8  \sqrt{\frac{ {2(1-p_{\alpha,s}^n)}}{{p_{\text{err}}}}}\right) \, \left(3+\frac{5}{4p_{\text{err}}}\right)^{-1},\\
 &\Delta (\eta)=4\log \frac{1}{\eta}+2\log \frac{2}{\zeta(\eta)^2}+2.  \label{lotsofdefinitions}
\end{align}
Furthermore,
the number of maximally entangled qubits that can be established  with infidelity at most $\epsilon$ through the remaining $n$ modes  is lower bounded by
\begin{equation}\label{eq:thm1a}
\sup_{\eta\in\left(0, \sqrt{\epsilon}-\epsilon'\right)} \left[n\log \frac{2\pi}{d^2}-2n \log \gamma(t+\mu (\zeta' (\eta)))-\Delta (\eta) +1\right],
\end{equation}
where $\zeta'$ is defined in the same way as  $\zeta$ except that  $\epsilon$ replaced by $2\epsilon$.  This bound can be achieved by  sending through each mode half of a two-mode squeezed state with average photon number $\bar n=  (10^{\frac{s}{10}}-1)/2$.  
\end{thm}
 The proof of the theorem is provided in the Supplemental Material \footnote{See the supplemental material, which contains Refs.~\cite{laurent2000,barnum2000,renner2008,konig2009,tomamichel2009,PhysRevA.71.062310,vitanov2013,morgan2013,kiukas2010,tomamichel2012,furrer2011,furrer2012,furrer2014,tomamichel2016,leverrier2017,khatri2020,wang2019,raju2020,RevModPhys.81.299}}.  
%This capacity lower bound is a function of $\epsilon$, $n$, $t$, $p_{\alpha}$ and $p_{\text{err}}$.
%Fig.~\ref{fig:scaling} shows the lower bounds on the regularized one-shot quantum capacity for different values of threshold $t$ in the correlation test. 
%The other parameters $k$, $n$, $t$, $p_{\text{err}}$ and $\epsilon$ are determined by Alice and Bob. 
 In Fig.~\ref{fig:scaling}, we show numerical plots of the bound~(\ref{eq:thm1}) for different values of $d$, $t$, $k/n$, and $\alpha$, setting  $\bar{n}=9.5$, corresponding to $13$ dB single-mode squeezing, achievable by state-of-the-art technology~\cite{vahlbruch2016}.
    The figure shows that the lower bound (\ref{eq:thm1}) can be raised  by increasing $k/n$,  and/or by reducing $d$ and/or by reducing $t$.   
%However, the lower bound does not increase much by further reducing $d$ 
%when the number of discretization intervals gets saturated. 
 %The lower bound is conditioned on the event of passing the correlation test. 
 Regarding the cut-off parameter $\alpha$,  it should be chosen to be large enough that the parameter $\lambda$ defined in Eq. (\ref{lotsofdefinitions}) does not exceed $\sqrt{\epsilon/2}$, for otherwise the bound  (\ref{eq:thm1}) on the  capacity  $Q^{\epsilon}$ becomes trivially 0.

 The probability of success of our protocol depends on the channel $\mathcal E$.  For example, if $\mathcal E$ is a pure loss channel, obtained by sending each input mode through a beamsplitter with transmissivity $\tau$, the success probability is approximately
$\frac{1}{2}+\frac{1}{2}\text{erf}\left[\sqrt{\frac{k}{\pi-2}}(\frac{td\sqrt{\pi}}{2(\sqrt{\bar{n}+1}-\sqrt{\tau\bar{n}})}-1)\right]$.

%Higher squeezing yields stronger correlation between input and output, and hence raises the probability to pass the correlation test. On the other hand, increase of product $td$ also raises the probability to pass the test.}

 %The main steps are as follows. 
%{\color{red} We consider the $2n$-partite joint state $\rho^{2n}$ generated by applying the channel locally on $n$ copies of input state, each of which is in a two-mode squeezed vacuum state with an external reference mode.
% Using results in~\cite{barnum2000,buscemi2010,morgan2013,tomamichel2016,pfister2018}, we lower bound $Q^\epsilon$ in terms of the conditional smooth max-entropy~\cite{renner2008,konig2009,furrer2011,berta2016} that is the maximum fidelity of $\rho^{2n}$ with a $n\times n$ partite product state that is completely mixed in the $n$ reference modes.
% Then,  our main technical contribution is to estimate the conditional smooth max-entropy of $\rho^{2n}$.}
 %to the estimation of the smooth max-entropy of a classical-quantum state obtained by performing homodyne measurements on the $n$ reference modes and discretizing the outcomes, which  can be further bounded if a suitable  correlation test is passed.  

{\em Protocol for independent and identical noises.}
%{\color{red}     Since relying on uncertainty relation~\cite{furrer2011,furrer2014}, the above protocol requires state preparations and measurements in two conjugate bases, i.e.\ position and momentum. Here we introduce a protocol using coherent states and heterodyne detections that is measurement in the basis of coherent states.}
 The previous  protocol can be applied to all correlated noisy quantum channels. However,  for some important i.i.d noisy channels, the lower bound in Eq.~(\ref{eq:thm1}) can be far from the optimal asymptotic lower bounds known in the literature~\cite{RevModPhys.86.1203}.
To address this problem, we now introduce another protocol that works specifically for i.i.d. channels.
Besides providing a better lower bound, our second protocol has the additional benefit that it does not require squeezing, but only the preparation of coherent states.   Since the protocol works in the coherent state basis, the homodyne detection in the first protocol   will be replaced by  heterodyne detection, which  is the canonical  measurement in the coherent state basis.

%Alice prepares $k$ coherent states, each of whose mean value $\beta_{A_i}\in\mathbb{C}$ is a random variables following a rotationally symmetric Gaussian distribution in the complex plane, with variance $2\bar{n}+3/2$. Then Alice multiplies each $\beta_{A_i}$ with a random number $u_i\in U(1)$, denoting the result as $\bm{x}:=(u_1\beta_{A_i}, \cdots, u_k\beta_{A_k})^\top \in\mathbb{C}^{k}$.
%At the output, Bob applies a single-mode heterodyne measurement on each of the $k$ modes, obtaining outcomes $\beta_{B_i}$. Then Bob multiplies $u_i$ with each corresponding outcome, obtaining $\bm{y}:=(u_1\beta_{B_i}, \cdots, u_k\beta_{B_k})^\top \in\mathbb{C}^{k}$.

The protocol works for phase-insensitive Gaussian channels, that is, Gaussian channels   $\Lambda$ satisfying the covariance condition ${\Lambda}  \circ  {\cal  U}_\theta  =  {\cal  U}_\theta \circ {\Lambda}$ for every $\theta \in  [0,2\pi]$ where ${\cal U}_\theta$ is the unitary channel corresponding to the operator $U_\theta  =  \exp  [- i \theta a^\dag a]$. This class of channels includes important examples in quantum optics and quantum communication, such as  optimal parametric amplifiers~\cite{cerf2000,chiribella2013}, Gaussian additive channels, and Gaussian loss channels~\cite{weedbrook2012}.  

In the protocol, Alice prepares $k$ coherent states, whose mean values $\bm{x}\in\mathbb{C}^{k}$ are random variables following a rotationally symmetric Gaussian distribution in the complex plane, with variance  equal $2\bar{n}+3/2$.
At the output,  Bob applies a single-mode heterodyne measurement on each of the $k$ modes, obtaining outcomes $\bm{y}\in\mathbb{C}^{k}$. 

 At this point, Alice and Bob can test the correlations between $\bm x$ and $\bm y$, as well as the amount of noise added by the channel.  Specifically, they can estimate the variance of  Bob's outcomes $\bm y$  and their cross-correlation with Alice's inputs $\bm x$. The result of the estimates are two values in a suitable confidence intervals, which contain the true values with probability $1-\delta$.   Here, the parameter $\delta$ can be chosen by Alice and Bob depending on how reliable they want their test to be. The results of the estimate are then used to infer a bound on the quantum capacity. The intuition is that higher cross-correlations and lower added-noise witness higher values of the capacity.    To make this intuition rigorous, we consider the minimum value of the cross-correlation and  the largest value of the added-noise in their respective confidence intervals. These two values, denoted by  $\gamma_{\min}$ and $\sigma_{\max}$, are given by 
 \begin{align}
   %& \gamma_A:= \frac{1}{2k}\left( 1+2\sqrt{\frac{\log(72/\delta)}{k}}\right)|| \bm{x}||^2-1, \\
   \sigma_{\max}:=&\frac{ ||\bm{y}||^2}{2(k-\sqrt{2k\ln1/\delta})}-1/2, \\
\nonumber   \gamma_{\min} :=&\frac{||\bm{x}||^2+||\bm{y}||^2+2\braket{\bm{x},\bm{y}}}{4(k+\sqrt{2k\ln 2/\delta}+\ln2/\delta)}-\bar{n}\\
  &- \frac{ ||\bm{y}||^2}{4(k-\sqrt{2k\ln1/\delta})}-3/4.
\end{align}
The conditions of high cross-correlation and low added-noise are then expressed as $\gamma_{\min}  \ge c$ and $\sigma_{\max} \le a$, respectively, where $c$  and $a$ are suitable thresholds that can be adjusted by Alice and Bob in the data analysis phase. The only constraint on $c$ and $a$ is that they need to be compatible with a quantum state, that is, that the matrix 
\begin{align}\label{eq:cvmatrix}
\xi  :=   \begin{pmatrix}
     (2\bar{n}+1)\mathds{1} & c\, \sigma_z \\
     c   \,  \sigma_z & a \mathds{1}
     \end{pmatrix}   \end{align}
satisfies the {\em bone fide} conditions for the covariance matrix of a quantum state  \cite{serafini2006}.
%The thresholds \color{red} $\Sigma_a^{\max}$, $a$ and $c$ can be used to obtain the covariance matrix~(\ref{eq:cvmatrix}) characterizing a Gaussian state at the boundary of a confidence region of two-mode quantum states, with confidence $1-\delta$.

%For well-studied Gaussian phase-insensitive channels~\cite{RevModPhys.84.621}, a random unitary operation is unnecessary as the outputs have rotational symmetry on phase space.
%However, in general, the i.i.d assumption can be broken by a global random unitary operation, in which case we suppose the noisy channels are covariant with respect to this postselection operation, similar to the assumptions in Refs.~\cite{leverrier2017,leverrier2019}.
% Then we have the following theorem.

%We suppose the noisy channel is a Gaussian phase-insensitive channel, then we have the following theorem.

\begin{thm}
If both  conditions $\gamma_{\min}  \ge c$ and $\sigma_{\max} \le a$ are satisfied, then, 
with error rate no larger than $\delta$, 
 the one-shot quantum capacity of the channel $\Lambda^{\otimes n}$  is lower bounded by 
 \begin{equation}\label{eq:thm2}
    Q^{\epsilon} \ge  n\,  \max\left\{  0,   g(a)-g(\nu_1)-g(\nu_2) - \inf_{\eta\in\left (0, \sqrt{\epsilon/2}\right)} 
      \frac{ h  (\eta)}k \right\} ,
\end{equation}
where 
$g(x):=\frac{x+1}{2}\log\frac{x+1}{2}-\frac{x-1}{2}\log\frac{x-1}{2}$, 
$\nu_1$ and $\nu_2$ are the symplectic eigenvalues of the matrix $\xi$ in Eq. (\ref{eq:cvmatrix}), and %\begin{align}
$h(\eta)  :=  \omega  \,   \sqrt{\log [2/(\sqrt{\epsilon/2}-\eta)^2]}-4\log\eta+2$, 
%\end{align}
with $\omega     :  =4\sqrt{k}\log(2\sqrt{1+\bar{n}}+2\sqrt{\bar{n}}+1)$.
\end{thm}

\begin{figure}
    \centering
    \includegraphics[width=0.42\textwidth]{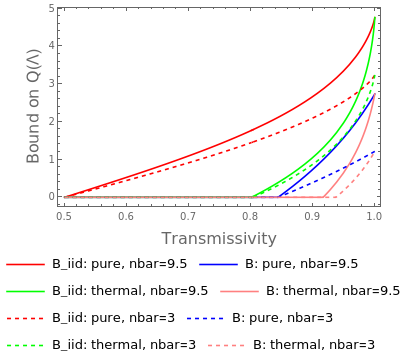}
    \caption{Asymptotic lower bounds $Q^{\rm iid}_{\rm asym}$ and $Q^{\rm general}_{\rm asym2}$ for i.i.d.  Gaussian  loss channels.    Here we show the case of {\em pure} loss (corresponding to $\bar{n}_{\text{th}}=0$) and of a {\em thermal} loss with   $\bar{n}_{\text{th}}=1$. The bounds are shown as functions of transmissivity $\tau$, for two values of  input photon number   $\bar{n}=9.5$ and $\bar{n}=3$.  For the non-iid protocol, we set the discretization parameter to  $d=0.1$.}   
    \label{fig:lossyChannel}
\end{figure}

 The proof of the theorem is given in the Supplemental Material. 
  When $n$ grows linearly with $k$,   Eq. (\ref{eq:thm2}) yields a lower bound on the asymptotic i.i.d. capacity $Q(\Lambda)  =  \lim_{\epsilon\to 0  }\lim_{n\to \infty} Q^{\epsilon}/n$, which reads
  \begin{equation}\label{asymiid}
  Q(\Lambda)  \ge    B_{\rm iid}:=\max \left\{  0,  g(a)-g(\nu_1)-g(\nu_2)\right\}.    
\end{equation} 
This asymptotic lower bound  can be compared with the analogous lower bound obtained from Eq. (\ref{eq:thm1}), which reads
\begin{equation}
   Q(\Lambda)  \ge  B:=\max  \left\{  0, \log   \frac{2\pi}{d^2}-2 \log\gamma(t)\right\},
\end{equation} where $\gamma(t)$ is the function defined in Eq. (\ref{lotsofdefinitions}). 
In Fig.~\ref{fig:lossyChannel}, we compare both asymptotic lower bounds for a practically important type of channels, namely Gaussian loss channels, corresponding to the transmission of the input through an arm of a beamsplitter with transmissivity $\tau$,  with a thermal state with mean photon number $\bar{n}_{\text{th}}$ in the other arm.    
For our comparison, we choose the threshold values that maximize the asymptotic bounds under the condition that the probability to pass the test approaches 1 in the asymptotic limit (see the Supplemental Material for the details of the optimization). In the i.i.d. case, the optimal thresholds turn out to be 
   $a=\tau(2\bar{n}+1)+(1-\tau)(2\bar{n}_{\text{th}}+1) $ and $c=\sqrt{\tau} \sqrt{(2\bar{n}+1)^2-1}$. 
  Inserting the optimal thresholds  into the expression of the asymptotic bound (\ref{asymiid}) for the Gaussian pure loss channel, we find the value  $B_{\rm iid}=\max \{ 0,  g((1-\tau)(2\bar{n}+1))-g(\tau(2\bar{n}+1))\}$. Remarkably,  this value is exactly equal to the energy-constrained quantum capacity of the channel~\cite{PhysRevA.63.032312,wilde2018}.

%{\em Protocol for qubit channels.}
%We also develop  a protocol
%using single-qubit preparations and measurements, 
%to estimate lower bounds on one-shot quantum capacity of qubit channels with i.i.d.\ noise, and its generalization into non-i.i.d.\ scenario by utilizing the exponential de Finetti theorem~\cite{renner2007,renner2008}, 
%Quantum AEP~\cite{tomamichel2009} implies that a lower bound on one-shot quantum capacity can be obtained from estimating coherent information. 
%To reliably estimate coherent information, we apply quantum process tomography, obtaining a confidence polytope~\cite{wang2019} of the Choi state. By minimizing the coherent information within this polytope, we obtain a lower bound on the one-shot quantum capacity.
%which approximates any permutation invariant state to an almost i.i.d.\ state. By relaxing the i.i.d.\ assumption in quantum AEP and quantum state tomography to almost i.i.d., we obtain a lower bound on one-shot quantum capacity with any arbitrary correlated noise, 
%as shown in the Supplemental Material.

{\em Conclusion.~}
We have introduced two protocols for  experimentally estimating lower bounds on  quantum capacities  of  CV channels in the realistic scenario where the channel under consideration is used a finite number of times.  
The first protocol applies to arbitrarily correlated, dynamically changing channels, possibly under the control of a malicious attacker, while the second protocol is restricted to  i.i.d phase insensitive Gaussian channels, and has a simpler experimental implementation.  
Both protocols can be implemented using current technologies on optical platforms. They provide a flexible method to validate practical quantum communication devices and quantum memories. In the longer term, they could be employed to discover useful quantum communication channels in quantum networks where the behavior of the transmission lines changes dynamically or adversarially.  Similarly, they could be used  witness the presence of  causal relations between quantum systems and to estimate the amount of quantum coherence between causally connected systems~\cite{maclean2017,bai2021}.  

{\em Acknowledgement.~}
We thank Chiara Macchiavello, Massimiliano F.~Sacchi, Quntao Zhuang, Zheshen Zhang, Nana Liu, Kunal Sharma, Ge Bai, Yan Zhu and Yuxiang Yang for the stimulating discussions.
 YDW and GC acknowledge funding from the Hong Kong Research Grant Council through grants no.\ 17300918 and no.\ 17307520, though the Senior Research Fellowship Scheme SRFS2021-7S02, the Croucher Foundation, and the John Templeton Foundation through grant 61466, The Quantum Information Structure of Spacetime (qiss.fr). Research at the Perimeter Institute is supported by the Government of Canada through the Department of Innovation, Science and Economic Development Canada and by the Province of Ontario through the Ministry of Research, Innovation and Science. The opinions expressed in this publication are those of the authors and do not necessarily reflect the views of the John Templeton Foundation.

\bibliography{refs}

\begin{widetext}

\section{More numerical analysis}

In this section, we present more numerical simulation results to analyze the dependence of the inferred lower bound of quantum capacity on those adjustable parameters in the protocols. We first study how the cutoff threshold $\alpha$ and the expected photon number $\bar{n}$ in the first protocol affect the lower bound in Theorem~1. As shown by Fig.~\ref{fig:cutoff}, when we fix $\bar{n}$ while reduce the value of $\alpha$, the protocol can fail to provide a nontrivial lower bound on the quantum capacity as $n$ keeps increasing. This is because for any given $p_{\alpha,s}$, there is an upper bound on adjustable $n$, above which, the lower bound in Theorem~1 doesn't work due to the fact that $\sqrt{\epsilon/2}-\lambda\le 0$ and the set of adjustable $\eta$ becomes empty.
On the other hand, if we increase $\alpha$, the lower bound is reduced in the region of positive bound. 
As shown by Fig.~\ref{fig:photonNum1}, when we fix $\alpha$ and increase $\bar{n}$, the protocol can fail to provide a nonzero bound as $n$ increases. This is because of the same reason as we discussed above.

\begin{figure}[h]
\begin{center}
\includegraphics[width=0.38\textwidth]{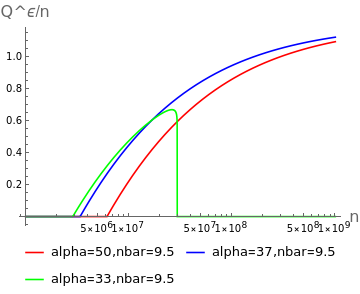}
\caption{Lower bound on $\frac{Q^{\epsilon}}{n}$, with $\epsilon=0.02$, given by Theorem~1, as a functions of~$n$ for different cutoff values $\alpha=50$, $\alpha=37$ and $\alpha=33$, respectively.
Other parameters are $d=0.1$, $t=3$, $k=n$ and $p_{\text{err}}=0.5$, and $\bar{n}=9.5$.
}
\label{fig:cutoff}
\end{center}
\end{figure}

\begin{figure}[h]
\begin{center}
\includegraphics[width=0.38\textwidth]{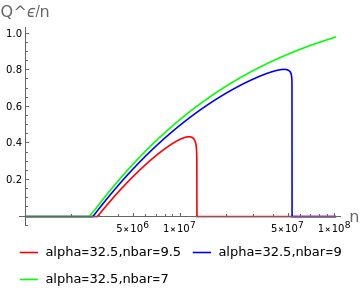}
\caption{Lower bound on $\frac{Q^{\epsilon}}{n}$, with $\epsilon=0.02$, given by Theorem~1, as a function of~$n$ for different expected photon number $\bar{n}=9.5$, $\bar{n}=9$ and $\bar{n}=7$, respectively.
Other parameters are $d$=0.1, $t=3$, $k=n$ and $p_{\text{err}}=0.5$, $\alpha=32.5$.
}
\label{fig:photonNum1}
\end{center}
\end{figure}

\begin{figure}[h]
\begin{center}
\includegraphics[width=0.38\textwidth]{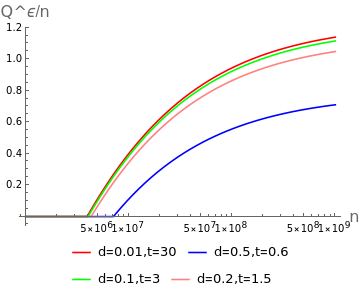}
\caption{Lower bound on $\frac{Q^{\epsilon}}{n}$, with $\epsilon=0.02$, given by Theorem~1, as a function of~$n$ for different combinations of discretization widths $d$, while $dt=0.3$ is kept.
Other parameters are $k=n$, $p_{\text{err}}=0.1$, $\alpha=37$, and $\bar{n}=9.5$.
}
\label{fig:binWidth}
\end{center}
\end{figure}

\begin{figure}[h]
\begin{center}
\includegraphics[width=0.45\textwidth]{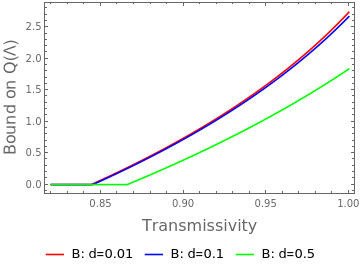}
\caption{Asymptotic lower bound $B$ for Gaussian pure loss channel as a function of transmissivity~$\tau$ for different values of discretization width $d$ and threshold $t=\frac{1}{d}\sqrt{\frac{4}{\pi}}\sqrt{\bar{n}(1+\tau)+1-2\sqrt{\bar{n}(\bar{n}+1)\tau}}$.
}
\label{fig:binWidthAsym}
\end{center}
\end{figure}

Then we investigate how different combinations of $d$ and $t$ affect the inferred lower bound while keeping $dt$ fixed. By reducing $d$ and fixing $dt$, we increase the number of bins within a fixed region of real numbers. As shown by Fig.~\ref{fig:binWidth}, reducing $d$ from $0.5$ to $0.1$, while keeping $dt=0.3$, raises the lower bound significantly. However, when we further reduce $d$ from $0.1$ to $0.01$, there is only a tiny increase in the lower bound. This phenomenon is reasonable because when we reduce $d$ from $0.5$ to $0.1$, the estimation of correlation between input and output becomes more accurate, which statistically yields more information about the quantum channel under study. Given a fixed passing region given by $dt$, this more information leads to an increase of the inferred lower bound on $Q^\epsilon$. However, when we further reduce $d$, $d$ no longer dominate the change of the lower bound. 
Similar phenomenons are shown by Fig.~\ref{fig:binWidthAsym} in the asymptotic limit.

Last, we study how the lower bound of $Q^\epsilon$ depends on the tolerable infidelity $\epsilon$. As Fig.~\ref{fig:infidelity} suggests, increasing $\epsilon$ raises the lower bound, but this change is not quite significant.

\begin{figure}[h]
\begin{center}
\includegraphics[width=0.38\textwidth]{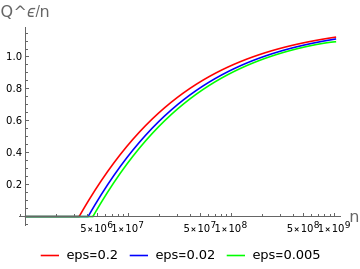}
\caption{Lower bounds on $\frac{Q^{\epsilon}}{n}$, given by Theorem~1, as functions of~$n$ for different maximal tolerable infidelities $\epsilon=0.2$, $\epsilon=0.02$ and $\epsilon=0.005$. 
Other parameters are $d=0.1$, $t=3$, $k=n$ and $p_{\text{err}}=0.1$, $\alpha=37$, and $\bar{n}=9.5$.
}
\label{fig:infidelity}
\end{center}
\end{figure}

%In the second protocol, besides thresholds $a$ and $c$, there is one more parameter $\delta$. $\delta$ characterizes the statistical errors in  $\sigma_{\max}$ and $\gamma_{\min}$ that are the estimators of a covariance matrix of a two-mode state, obtained by apply the quantum channel to an input mode, which is in a two-mode squeezed vacuum state with an additional reference mode. $\delta$ does not affect the lower bound on $Q^\epsilon$, but $\delta$ affects the probability to pass the . 

\section{Further discussion on parameter choices}
 The adjustable parameters are either related to the limitation of the
experimental setups available to Alice and Bob (e.g. limits on the maximum energy of the
input state, limits on the resolution of the detectors, limits to the amount of squeezing in the
input states), or related to the required degree of confidence in their estimation procedure.  We will discuss the meanings of each parameter and how to choose those parameters in detail in the following.

In the first protocol, $d$, $\alpha$ and $\bar{n}$ depend on the practical quantum devices in the protocol. Specifically, $d$ represents the resolution of homodyne detections associated to the width of detector pixels. In our numerical calculations, we choose $d$ as $0.01$, $0.1$, and $0.5$. 
$(-\alpha, \alpha)$ is the maximal range of displacement that can be performed by Alice’s quantum device. 
$\alpha$ must be large enough to make $\sqrt{\epsilon/2}-\lambda> 0$, otherwise, the bound in Theorem 1 fails to give a nontrivial value.
$\bar{n}$ denotes the expected photon number per input mode, which is limited by achievable squeezing. Any $\bar{n}\le 13.6$ is achievable by current technology~\cite{vahlbruch2016}, corresponding to squeezing below $15$ dB.
$p_{\alpha,s}=\text{erf}(\alpha/\sqrt{2(2\bar{n}+1)})$ is determined by $\alpha$ and $\bar{n}$, denoting the probability that a sample from zero-mean Gaussian distribution with variance $2\bar{n}+1$ falls within $(-\alpha, \alpha)$, where erf is called error function.

$k$ is the number of modes Alice and Bob can sacrifice for performing test and $n$ is the number of modes, which are demanded by Alice and Bob for later quantum communication. Both $k$ and $n$ are practically upper limited, and $k$, $n\le 10^9$ are considered to be within practical regime in CV QKD~\cite{leverrier2017}, whose setting is similar to ours. 
 The ratio $k/n$, and other parameters $t$, $p_{err}$ and $\epsilon$ can be chosen by users of the channel, Alice and Bob.  $t$ is the threshold value in the test. Reducing $t$  can increase the inferred lower bound on quantum capacity, but simultaneously reduce the probability to pass the test, and vice versa.  $0<p_{err}<1$ is the tolerable error probability in the inference. $\epsilon$ is introduced in the definition of one-shot quantum capacity and denotes the tolerable infidelity of quantum communication. In the main text, we choose $\epsilon=0.02$, which is acceptable considering the fact that experimental fidelity of qubit teleportation over long distance is just above~$0.9$~\cite{raju2020}.

In the second protocol, $k$, $n$ and $\epsilon$ have the same meanings as those we have discussed in the first protocol.
%For a prior unknown i.i.d noisy quantum channel, to choose threshold parameters $a$ and $c$, Alice and Bob may presuppose the noisy channel is a Gaussian thermal lossy channel with minimal tolerable transmissivity $0<\tau<1$ and maximal tolerable thermal noise $2\bar{n}_{\text{th}}+1$. Then by noting that $\sigma_{\max}$ and $\gamma_{\min}$ are estimators of the parameters in the cavariance matrix of two-mode state, obtained by applying the noisy channel to one party of a two-mode squeezed state, Alice and Bob can choose $a=\tau_{\min}(2\bar{n}+1)+(1-\tau_{\min})v_{\max} $ and $c=\sqrt{\tau_{\min}} \sqrt{(2\bar{n}+1)^2-1}$. 
%However, this choice may yield a lower probability to pass the test or a trivial lower bound, if the i.i.d noisy channel is too far from a Gaussian channel. 
The threshold parameter $a$ represents maximal tolerable additional noise in $\bm{y}$ and the other threshold parameter $c$ represents minimal tolerable cross-correlation between $\bm{x}$ and $\bm{y}$. $a$ and $c$
should be chosen to make matrix
$\xi=
\begin{pmatrix}
(2\bar{n}+1)\mathds{1} & c\sigma_z \\
c\sigma_z  & a\mathds{1}
\end{pmatrix}
$
a viable covariance matrix, that is to satisfy the bona fide conditions~\cite{serafini2006}
\begin{equation*}
   \xi+\text{i}\Omega\ge 0, \text{ where } \Omega:= \begin{pmatrix}
    0 & 1 & 0 & 0\\
    -1 & 0 & 0 & 0\\
    0 & 0& 0 & 1\\
    0 & 0 & -1 & 0
   \end{pmatrix}.
\end{equation*}
Suitable $a$ and $c$ can be adjusted by Alice and Bob in the data analysis phase after obtaining measurement outcomes.
 Basically, Alice and Bob can pick the values that give them the best bound on the capacity of the channel under consideration.  The optimal values of $a$ and $b$ generally depend on the channel itself. 
%In Fig.2 (b), to choose $a$ and $c$, we presuppose the i.i.d noisy channel is a Gaussian thermal loss channel with minimal tolerable transmissivity being $0.75$ and maximal tolerable additional noise being vacuum noise. Then we choose $a=0.75(2\bar{n}+1)+0.25=15.25$ and $c=\sqrt{0.75} \sqrt{(2\bar{n}+1)^2-1}=17.3$. We also consider the case, where $a$ is slightly higher and $c$ is slightly lower, respectively. 
%Another choosable parameter is $\delta$, which characterizes the statistical error in estimators $\sigma_{\max}$ and $\gamma_{\min}$. Higher error probability $\delta$ reduces $\sigma_{\max}$ and raises $\gamma_{\min}$, and hence raises the probability to pass $\sigma_{\max}\le a$ and $\gamma_{\min}\ge c$.

Table~\ref{tab:my_label} lists the physical meanings of all the adjustable parameters in both protocols.

\begin{table}[]
    \centering
    \begin{tabular}{c|c}
      Parameter  & physical/statistical meaning \\\hline
       $d$ &  resolution of homodyne detections associated to width of detector pixels \\ 
       $\alpha$ & maximal range of displacement/ maximal range of reliable readout of homodyne detectors \\
       $\bar{n}$ &  expected photon number per input mode \\ 
       $n$ & number of modes for communication\\
       $k$ &  number of modes for test \\
       $t$ &  threshold value in test \\
       $p_{err}$ & tolerable error probability in the inference \\
       $\epsilon$ &  tolerable infidelity of quantum communication \\
       $a$ & threshold on maximal additional noise in $\bm{y}$\\
       $c$ & threshold on minimal correlation between $\bm{x}$ and $\bm{y}$\\
       $\delta$ &  error probability of the confidence intervals characterized by $\sigma_{\max}$ and $\gamma_{\min}$ \\
    \end{tabular}
    \caption{Physical meanings of all adjustable parameters in both protocols}
    \label{tab:my_label}
\end{table}

\section{Asymptotic limit for Gaussian loss channels}
In this section, we explain how to obtain the asymptotic limits of capacity bounds in Theorem~1 and Theorem~2 for Gaussian loss channels.
We first show how we obtain the lower bound of quantum capacity in the first protocol for i.i.d Gaussian loss channels in the asymptotic limit.
The entangled state at input mode and reference mode is a two-mode squeezed vacuum state $\ket{\Psi_{\rho_{\text{th}(\bar{k})}}}:=e^{\kappa/2(\hat{a}\hat{b}-\hat{a}^\dagger\hat{b}^\dagger)}\ket{0}\ket{0}$.
In Heisenberg picture, the position operators at input mode and reference mode can be written as
$\hat{q}_{A'}=\cosh \kappa \hat{q}_1^{(0)}+\sinh\kappa \hat{q}_2^{(0)}$ and $\hat{q}_{A}=\sinh \kappa \hat{q}_1^{(0)}+\cosh\kappa \hat{q}_2^{(0)}$, where $\hat{q}^{(0)}$ denotes the position operator of a vacuum state.
%It is easy to see that $q_{A'}-q_{A}\sim \mathcal N(0, \sqrt{2}\text{e}^{-\kappa})$ and $p_{A'}+p_{A}\sim \mathcal N(0, \sqrt{2}\text{e}^{-\kappa})$, where $\mathcal N(x, y)$ represents a Gaussian distribution with mean value~$x$ and standard deviation $y$. 
For a Gaussian loss channel with transmissivity $\tau$ and mean photon number of thermal noise $\bar{n}_{\text{th}}$, the position and momentum operators at output become $\hat{q}_B=\sqrt{\tau}\hat{q}_{A'}+\sqrt{1-\tau}\hat{q}_{\text{th}}$ and $\hat{p}_B=\sqrt{\tau}\hat{p}_{A'}+\sqrt{1-\tau}\hat{p}_{\text{th}}$. Hence
\begin{equation}
    \hat{q}_A-\hat{q}_B=(\sinh \kappa -\sqrt{\tau}\cosh \kappa) \hat{q}_1^{(0)}+(\cosh\kappa -\sqrt{\tau}\sinh\kappa) \hat{q}_2^{(0)}-\sqrt{1-\tau}\hat{q}_{\text{th}}.
\end{equation}
The random variable $q_A-q_B$ follows a Gaussian distribution with zero mean and standard deviation 
 $$\sqrt{(\sinh \kappa -\sqrt{\tau}\cosh \kappa)^2+(\cosh\kappa -\sqrt{\tau}\sinh\kappa)^2+(1-\tau)(2\bar{n}_{\text{th}}+1)}.$$ Then $|q_A-q_B|$ simply follows a half-normal distribution with mean value 
 $$
 \sqrt{2/\pi}\sqrt{(\sinh \kappa -\sqrt{\tau}\cosh \kappa)^2+(\cosh\kappa -\sqrt{\tau}\sinh\kappa)^2+(1-\tau)(2\bar{n}_{\text{th}}+1)}.
 $$
 $|p_A+p_B|$ follows the same distribution.
 From central limit theorem, the sample mean
 $\frac{1}{k}\sum_{i=1}^{k} |x_{A,i}- x_{B,i}|$
 approximately follows a Gaussian distribution with mean value
 $$
 \sqrt{2/\pi}\sqrt{(\sinh \kappa -\sqrt{\tau}\cosh \kappa)^2+(\cosh\kappa -\sqrt{\tau}\sinh\kappa)^2+(1-\tau)(2\bar{n}_{\text{th}}+1)}
 $$
 and standard deviation
 $$
 \sqrt{\frac{1-2/\pi}{k}}\sqrt{(\sinh \kappa -\sqrt{\tau}\cosh \kappa)^2+(\cosh\kappa -\sqrt{\tau}\sinh\kappa)^2+(1-\tau)(2\bar{n}_{\text{th}}+1)}.
 $$

When the number of channels uses $k$ is asymptotically large, averaged distance $1/k\sum_{i=1}^k |x_{A,i}-x_{B, i}|$ becomes a sharp distribution at its mean value. 
Thus, in the limit of asymptotic large number of uses, we can set 
\begin{align}\notag
t=&\frac{\sqrt{2/\pi}}{d}\sqrt{(\sinh \kappa -\sqrt{\tau}\cosh \kappa)^2+(\cosh\kappa -\sqrt{\tau}\sinh\kappa)^2+(1-\tau)(2\bar{n}_{\text{th}}+1)}\\
=&\frac{1}{d}\sqrt{\frac{4}{\pi}}\sqrt{\bar{n}(1+\tau)+1+\bar{n}_{\text{th}}(1-\tau)-2\sqrt{\bar{n}(\bar{n}+1)\tau}}
\end{align}
and the correlation test can almost always be passed, where the second line comes froms the fact $\sinh^2\kappa=\bar{n}$.

Then we explain how we obtain the asymptotic lower bound of quantum capacity in the second protocol for i.i.d Gaussian loss channels.
We consider a two-mode Gaussian state obtained by applying a Gaussian loss channel, with transmissivity $\tau$ and thermal mean photon number $\bar{n}_{\text{th}}$, on half of a two-mode squeezed vacuum state with covariance matrix 
\begin{equation*}
\begin{pmatrix}
(2\bar{n}+1)\mathds{1} & \sqrt{(2\bar{n}+1)^2-1}\sigma_z \\
\sqrt{(2\bar{n}+1)^2-1}\sigma_z & (2\bar{n}+1)\mathds{1}
\end{pmatrix}.
\end{equation*}
Simple calculation indicates that this resulting Gaussian state has covariance matrix
 \begin{equation*}
     \begin{pmatrix}
 2(\bar{n}+1) \mathds{1} & \sqrt{\tau}\sqrt{(2\bar{n}+1)^2-1}\sigma_z \\
 \sqrt{\tau}\sqrt{(2\bar{n}+1)^2-1}\sigma_z & \left[\tau(2\bar{n}+1)+(1-\tau)(2\bar{n}_{\text{th}}+1)\right]\mathds{1}
 \end{pmatrix}.
 \end{equation*}
 The definition of $\sigma_{\max}$ and $\gamma_{\min}$ indicates that, when $k\rightarrow \infty$, $\sigma_{\max}$ is an unbiased estimator of variance $\tau(2\bar{n}+1)+(1-\tau)(2\bar{n}_{\text{th}}+1)$ and $\gamma_{\min}$ is an unbiased estimator of covariance $\sqrt{\tau}\sqrt{(2\bar{n}+1)^2-1}$.
As the statistical errors from finite sampling in estimators $\sigma_{\max}$ and $\gamma_{\min}$ asymptotically go to zero, we choose 
$a=\tau(2\bar{n}+1)+(1-\tau)(2\bar{n}_{\text{th}}+1)$ and $c=\sqrt{\tau}\sqrt{(2\bar{n}+1)^2-1}$, which guarantees that the probability of pass asymptotically approaches unit. 
 For Gaussian pure loss channel, 
 \begin{align}\notag
     B_{iid}&=g(\tau(2\bar{n}+1)+1-\tau)-g(\nu_1)-g(\nu_2) \\
     & =g((1-\tau)(2\bar{n}+1))-g(\tau(2\bar{n}+1)),
 \end{align}
 where $\nu_1$ and $\nu_2$ are the symplectic eigenvalues of  \begin{equation*}
     \begin{pmatrix}
 2(\bar{n}+1) \mathds{1} & \sqrt{\tau}\sqrt{(2\bar{n}+1)^2-1}\sigma_z \\
 \sqrt{\tau}\sqrt{(2\bar{n}+1)^2-1}\sigma_z & \left[\tau(2\bar{n}+1)+1-\tau\right]\mathds{1}
 \end{pmatrix}.
 \end{equation*}
 This equals to the energy-constrained asymptotic quantum capacity with mean photon number $\bar{n}$.

\section{Proof of Theorem 1}
\label{sec:proofThm1}
The one-shot quantum capacity  is defined as
\begin{equation}
Q^{\epsilon}(\mathcal E):=\max\{\log b | F(\mathcal E, b) \ge1-\epsilon\},
\end{equation}
where $b$ is the dimension of the subspace in which information is encoded, and
\begin{equation}
F(\mathcal E, b):=\max_{\bar{\mathcal H}\subset \mathcal H, \text{dim}(\bar{\mathcal H})=b} \max_{\mathcal D}\min_{\ket{\phi}\in\bar{\mathcal H}}\braket{\phi|\mathcal D\circ \mathcal E(\ket{\phi}\bra{\phi})|\phi},
\end{equation}
is  the maximum  fidelity obtained by optimizing the choice of encoding subspace $\overline {\cal H}$ and the choice of a decoding channel $\cal D$, in the worst case over all possible input states. 
    When the channel is of the form $\mathcal E=\Lambda^{\otimes n}$, corresponding to $n$ i.i.d.  uses of a channel $\Lambda$, the asymptotic quantum capacity $Q(\Lambda)$ is equal to the limit of  the regularized one-shot  capacity $Q^{\epsilon}(\Lambda^{\otimes n})/n$   when  the number of uses goes to infinity and the error tolerance goes to zero.
%\begin{equation}
%Q(\Lambda)=\lim_{\epsilon\rightarrow 0}\lim_{n\rightarrow \infty}Q^{\epsilon} (\Lambda^{\otimes n})/n \,.
%\end{equation}
In summary, the one-shot quantum capacity includes as a special case the asymptotic quantum capacity. 

We then present all the related concepts of min- and max-quantum entropies~\cite{renner2008,konig2009}, which are rigorously generalized into infinite dimensions~\cite{furrer2011}.
The min-entropy of $\rho_{AB}$ given $\sigma_B$ is
\begin{equation}
    H_{\min}(\rho_{AB}|\sigma_B):=-\log_2 \min\{\lambda|\lambda\mathds{1}\otimes \sigma_B\ge \rho_{AB}\},
\end{equation}
and the min-entropy of $\rho_{AB}$ given system B is
\begin{equation}
    H_{\min}(A|B)_{\rho}:=\sup_{\sigma_B}H_{\min}(\rho_{AB}|\sigma_B).
\end{equation}
Given a purification $\rho_{ABC}$ of $\rho_{AB}$, the max-entropy of $\rho_{AB}$ given system $B$ is
\begin{equation}
    H_{\max}(A|B)_{\rho_{AB}}:=-H_{\min}(A|C)_{\rho_{AC}}.
\end{equation}

Similarly, one can define the smooth min-entropy 
\begin{equation}
    H_{\min}^\epsilon(\rho_{AB}|\sigma_B):= \max_{\rho'_{AB}\in B^\epsilon(\rho_{AB})}H_{\min}(\rho'_{AB}|\sigma_B),
\end{equation}
where $B^\epsilon(\rho):=\{\rho'\ge 0| \tr\rho'\le 1, \mathcal P(\rho, \rho')\le \epsilon\}$ is an $\epsilon$-ball around~$\rho$ with
 $\mathcal P(\rho, \rho'):=\sqrt{1-|| \sqrt{\rho}\sqrt{\rho'}||^2_1}$ called purified distance,
and
\begin{equation}\label{def:smoothmin}
    H_{\min}^\epsilon(A|B)_\rho:=\max_{\rho'\in B^\epsilon(\rho)} H_{\min}(A|B)_{\rho'}.
\end{equation}
Given a purification $\rho_{ABC}$ of $\rho_{AB}$,  the smooth max-entropy of $\rho_{AB}$ is
\begin{equation}\label{eq:dualitySmoothEntropy}
    H_{\max}^\epsilon(A|B)_{\rho_{AB}}:=-H_{\min}^\epsilon(A|C)_{\rho_{AC}}.
\end{equation}

Suppose we apply a channel $\mathcal E: \mathcal H_{A'}^{\otimes n}\rightarrow \mathcal H_{B}^{\otimes n}$ to an input state $\sigma_{A'^n}$, where
$n$ denotes the number of subsystems.  
The purification of $\sigma_{A'^n}$ is $\ket{\Psi_\sigma}_{A'^n A^n}$. Then the joint state at reference $A^n$ and output $B^n$ is $\rho_{A^n B^n}:=\mathds{1}\otimes\mathcal E(\ket{\Psi_\sigma}\bra{\Psi_\sigma})$.

\iffalse
The idea to prove Theorem~1 is to reduce the estimation of a lower bound to the one-shot capacity to the estimation of an upper bound on  the smooth max-entropy of a  quantum state $\rho_{A^n B^n}$, obtained by feeding half of a two-mode squeezed state in each of the $n$ communication modes.    In turn, the estimation of the smooth max entropy of the quantum state $\rho_{A^n B^n}$ can be reduced to the estimation of  the smooth max-entropy of a classical-quantum state $\omega_{X^n B^n}$, where $\omega_{X^n B^n}$ is the joint post-measurement state conditioned on the previous test is passed, and $X^n$ denotes classical registers storing Alice's discretized outcomes of homodyne measurements.
 Our key result is the following bound 
\begin{equation}\label{eq:quantumMaxEntropyToClassical}
    H_{\max}^{3\chi+5\chi'}(A^n|B^n)_{\rho} \le  n\log_2 \frac{d^2}{2\pi}+2H_{\max}^{\chi'}(X^n|B^n)_\omega 
  -2\log_2 \frac{2}{\chi^2}
\end{equation}
for $d\ll 1$, and any $\chi, \chi'>0$. 
%The derivation of the bound is provided in the Supplemental Material~\footnote{See the supplemental material, which contains Refs.~\cite{vitanov2013,kiukas2010,tomamichel2012,khatri2020}.}.  
The strategy is to use a  CV entropic uncertainty relation derived in~\cite{furrer2012,furrer2014}, and adapt the result to practical homodyne measurements with a cutoff on the maximum values of the measured quadratures.  
$H_{\max}^{\chi'}(X^n|B^n)_{\omega}$ can be further bounded using $t$, if a suitable correlation test is passed.
\fi

\begin{lemma}[lower bound on one-shot quantum capacity as optimization of max-entropy~\cite{barnum2000,buscemi2010,morgan2013,tomamichel2016,pfister2018}]\label{lemma:capaitytoEntropy}
Given a quantum channel $\mathcal E$ from $\mathcal H_{A'}$ to $\mathcal H_B$,
the one-shot quantum capacity of $\mathcal E$ is bounded by
\begin{equation}
Q^{\epsilon}(\mathcal E)\ge \sup_{\eta\in \left(0,\sqrt{\epsilon/2}\right)} \max_{\sigma\in\mathcal S(\mathcal H_{A'}^{\otimes n})}\left(-H_{\max}^{\sqrt{\epsilon/2}-\eta}(A^n|B^n)_{\rho} +4\log_2 \eta\right)-2.
\end{equation}
\end{lemma}

We can drop the maximization over all possible input states by choosing a specific input $\sigma_{A'}$.
For infinite-dimensional quantum system, we can further restrict the energy of each input mode to obtain a lower bound on the energy-constrained one-shot quantum capacity.
In the following, we choose the input at each mode as a thermal state with mean photon number $\bar{n}$, i.e. $\rho_{\text{th}}(\bar{n})=\sum_{n=0}^{\infty}\frac{\bar{n}^n}{(\bar{n}+1)^{n+1}}\ket{n}\bra{n}$, whose purification is a two-mode squeezed vacuum state $\ket{\Psi_{\rho_{\text{th}(\bar{n})}}}:=e^{\kappa/2(\hat{a}\hat{b}-\hat{a}^\dagger\hat{b}^\dagger)}\ket{0}\ket{0}$ with $\cosh(2\kappa)=2\bar{n}+1$.

Below we present a lower bound, closely related to the above bound, on the maximal number of maximally entangled pairs, which can be established by applying entanglement distillation on $\rho_{A^n B^n}$.
\begin{lemma}[lower bound on distillable entanglement~\cite{morgan2013,tomamichel2016,khatri2020}]
\label{lemma:distilledEntanglement}
For any state $\rho_{A^n B^n}$, a lower bound of its one-shot distillable entanglement is
\begin{equation}
    \sup_{\eta\in \left(0,\sqrt{\epsilon}\right)} \left(-H_{\max}^{\sqrt{\epsilon}-\eta}(A^n|B^n)_{\rho} +4\log_2 \eta\right)-1.
\end{equation}
\end{lemma}
This Lemma shows that by estimating an upper bound of $H_{\max}(A^n|B^n)_{\rho}$, we can not only detect a lower bound on one-shot quantum capacity, but also obtain a lower bound on the amount of entanglement, which can be established by sending just halves of two-mode squeezed vacuum states. 

Hence, prediction of a lower bound on one-shot quantum capacity is now reduced to estimating smooth max-entropy of an unknown state resulting from the application of the channel to $n$ two-mode squeezed states. 
%   In the following, we will further simplify the probe, by  reducing the input to single-mode states. 
 An indirect way to  estimate $H_{\text{max}}^{\sqrt{\epsilon/2}-\eta}(A^n|B^n)_{\rho}$ would be to perform a full quantum tomography of the state $\rho_{A^n B^n}$~\cite{RevModPhys.81.299}.  However, full tomography is highly demanding for high-dimensional systems, and convergence issues from the use of finite statistics arise in the CV case.  
% In addition, the usual quantum tomography protocols only provide information on the original measured states, which are destroyed by the measurements, while gives no information about the unmeasured states, unless the latter are assumed to be identical to the measured ones~\cite{christandl2012}.
Moreover, even if we knew $\rho$ exactly, evaluating the smooth max-entropy by optimizing over a neighborhood of~$\rho$ is hard in general~\cite{renner2008}. To circumvent these problems, we  now propose a  method  to estimate an upper bound on the smooth max-entropy  without full tomography.  

Here we present the protocol for arbitrary unknown correlated noise in the entanglement-based formalism, instead of the one in the formalism of preparation and measurement shown in the main text.
Given a $(k+n)$-mode input and $(k+n)$-mode output channel, Alice prepares $k+n$ copies of two-mode entangled states $\ket{\psi}$ and feed one party of each to the channel.
Through negotiation, Alice and Bob agree on $k$ random pairs of modes. On these $k$ pairs, Alice and Bob both apply homodyne detections at each of them in the same random bases $\bm{z}^{k}\in \{0,1\}^{\otimes k}$ ($0$ dentoes position and $1$ denotes momentum).
Suppose the discretization distance when discretizing the outcomes is $d>0$ and the outcome cutoff is $(-\alpha+d, \alpha-d)$. Each measurement outcome is projected into one of the $2\alpha/d$ regions, $\{(-\infty,-\alpha+d], (-\alpha+d,-\alpha+2d], \dots, (\alpha-d,\infty)\}$. Accordingly each outcome is mapped to an integer in the set $ \chi:=\{0, 1, \dots, \frac{2\alpha}{d}-1\}$, where $d$ and $\alpha$ are chosen to make $2\alpha/d\in\mathbb{N}^+$.
$\bm{x}_A^{pe}\in \chi^{\otimes k}$ and $\bm{x}_B^{pe}\in \chi^{\otimes k}$ denote Alice's and Bob's discretized measurement outcomes at $k$ modes respectively.
Alice and Bob pass the test at the $k$ subsystems if the average distance
\begin{equation}
1/k\sum_{i=1}^{k} |x_{A,i}^{pe}- x_{B,i}^{pe}| \le t.
\end{equation}
Otherwise, they abort the protocol.

Denote the state at the other $n$ pairs of modes by $\rho_{A^nB^n}$, whose purification is denoted by $\rho_{A^n B^n E}$. 
Alice applies homodyne detections at the remaining $n$ modes on random chosen bases $z_n\in\{0,1\}^{\otimes n}$ and $\bm{x}_A\in \chi^{\otimes n}$ denotes Alice's measurement outcomes at these $n$ modes. 
Denote $\omega_{A^n X^n B^n}$ as the joint post-measurement state at $A^n$, $X^n$, $B^n$, conditioned on the previous test is passed, where $X^n$ denotes  classical registers storing Alice's discretized measurement outcomes~$\bm{x}_A$, and $\omega_{A^n X^n B^n E}$ as the purified state.

Now we present the proof of Theorem~1 by following the idea in \cite{pfister2018} and using mainly the technical tools proven in Ref.~\cite{furrer2012}.
Before we show the proof, we first present the following three useful lemmas.
\begin{lemma}[chain rule of smooth max-entropy]\label{lemma:chainrule}
Smooth max-entropy satisfies the following chain rule, for any $\epsilon>0$, $\epsilon',\epsilon''\ge 0$, and any $\sigma\in\mathcal S(\mathcal H_A\otimes \mathcal H_B\otimes \mathcal H_C)$, where $\mathcal H_A$, $\mathcal H_B$ and $\mathcal H_C$ can be infinite-dimensional Hilbert spaces,
\begin{equation}
H_{\max}^{\epsilon +\epsilon'+2\epsilon''} (AB|C)_\sigma \le H_{\max}^{\epsilon'} (A|BC)_\sigma + H_{\max}^{\epsilon''} (B|C)_\sigma+\log \frac{2}{\epsilon^2}.
\end{equation}
\end{lemma}

This lemma was first proven by Ref.~\cite{vitanov2013} for finite-dimensional state $\sigma$. This resulted can be extended to infinite-dimensional quantum system by combining the fact that max-entropy on infinite-dimensional Hilbert spaces can be asymptoticly approached by max-entropy on finite-dimensional Hilbert spaces~\cite{furrer2011} and the chain rule of smooth max-entropy in Ref.~\cite{vitanov2013}.

\begin{lemma}[CV entropic uncertainty relation~\cite{furrer2012}]\label{lemma:CVuncertainty}
 The post-measurement state $\omega$, conditioned on the test at $n$ modes being passed, satisfies the following entropic
uncertainty relation
\begin{equation}
H_{\min}^{\epsilon+2\epsilon'}(X^n| E)_\omega \ge -n\log c(d) -H_{\max}^{\epsilon}(X^n| B^n)_\omega,
\end{equation}
where $c(d)=\frac{d^2}{2\pi}S_0^{(1)}\left(1, \frac{d^2}{4}\right)^2$, 
$\epsilon'=\sqrt{\frac{2(1-(1-p_\alpha)^n)}{p_{\operatorname{pass}}}}$, 
$p_{\operatorname{pass}}$ denotes the probability that the test is passed,
and $p_\alpha$ is an upper bound of the probability that each $x_A$ exceeds the region $(-\alpha, \alpha)$.
\end{lemma}
Here $S_0^{(1)}(\cdot,\cdot)$ denotes the radial prolate spheroidal wave function of the first kind~\cite{kiukas2010,furrer2014} and when $d\ll 1$, we have $c(d)\approx d^2/(2\pi)$. 
If Alice's state preparation can be trusted, then the states in her possession are just copies of thermal states. For a thermal state $\rho(\bar{n})$, the variances of both quadratures are $2\bar{n}+1$. We can obtain the value of $p_\alpha$ from error function. For example, when $\alpha=37$ and $\bar{n}=9.5$, $p_\alpha= 1-\text{erf}(6.17)\approx 1.11\times 10^{-16}$.

Estimating $H^{\sqrt{\epsilon/2}-\eta}_{\max}(A^n|B^n)_\rho$ can be reduced to the estimation of $H^{\zeta'}_{\max}(X^n|B^n)$. 
 At this point, the intuition is that if both Alice and Bob apply homodyne detections in the same basis at certain pairs of modes and their outcomes are highly correlated, then $H_{\max}^{\zeta'}(X^n|B^n)_{\omega}$ must be small, because $B^n$ contains much information about~$A^n$. This intuition was made rigorous in Ref.~\cite{furrer2012} as given in the following lemma, which showed that if a suitable  correlation test is passed, $H_{\max}^{\zeta'}(X^n|B^n)_{\omega}$ can be bounded using the data of homodyne outcomes.  

\begin{lemma}[upper bound on max-entropy~\cite{furrer2012}]\label{lemma:boundMaxEntropy}
Conditioned on that $1/k\sum_{i=1}^k |X_{A,i}^{pe}-X_{B,i}^{pe}|\le t$, the smooth max-entropy of Alice's measurement outcomes $\bm{x}_A$, given Bob's system $B^n$ and measurement basis choices $\bm{z}_n$,
is bounded by
\begin{equation}
H_{\max}^{\frac{\epsilon}{4p_{\text{pass}}}-\frac{2f(p_{\alpha},n)}{\sqrt{p_{\text{pass}}}}}(X^n| B^n)\le n\log \gamma\left(t+\mu_0(\epsilon)\right),
\end{equation}
where $\gamma(t):=(t+\sqrt{1+t^2})\left(\frac{t}{\sqrt{1+t^2}-1}\right)^t$, $\mu_0(\epsilon)=\frac{2\alpha}{d}\sqrt{\frac{(k+n)(k+1)}{nk^2}\log{\frac{1}{\epsilon/4-2f(p_\alpha, n)}}}$, and $f(p_\alpha, n):=\sqrt{2(1-(1-p_\alpha)^n)}$.
\end{lemma}

Now we are ready to present the result of prediction of lower bounds on quantum capacities over $n$-mode quantum channels with general correlated noises.

\begin{thm}
 If the measurement outcomes at the $k$ test modes pass the test: $1/k\sum_{i=1}^k |x_{A,i}^{pe}-x_{B,i}^{pe}|\le t$, then either 
  the probability to pass this test is lower than $p_{\text{pass}}$, or
 the  one-shot quantum capacity of the channel corresponding to the remaining $n$ modes is bounded  by
\begin{equation}
Q^{\epsilon}\ge\max\left\{0, \sup_{\eta\in\left(0, \sqrt{\epsilon/2}-8f(p_{\alpha}, n)\left(3+\frac{5}{4p_{\text{pass}}}-\frac{1}{\sqrt{p_{\text{pass}}}}\right)\right)} \left[n\log_2 \frac{2\pi}{d^2}-2n \log_2\gamma\left(t+\mu_0(\zeta)\right)-4\log_2\frac{1}{\eta}
-2\log_2 \frac{2}{\zeta^2}-2\right]\right\},
\end{equation}
where
 $ \zeta=\left(\sqrt{\epsilon/2}-\eta+\frac{8f(p_\alpha, n)}{\sqrt{p_{\text{pass}}}}\right)/\left(3+\frac{5}{4p_{\text{pass}}}\right)$,
and the number of maximally entangled pairs, which can established by sending halves of two-mode squeezed vacuum states, can be lower bounded by
\begin{equation}
    \sup_{\eta\in\left(0, \sqrt{\epsilon}-8f(p_{\alpha}, n)\left(3+\frac{5}{4p_{\text{pass}}}-\frac{1}{\sqrt{p_{\text{pass}}}}\right)\right)} \left[n\log_2 \frac{2\pi}{d^2}-2n \log_2\gamma\left(t+\mu_0(\zeta')\right)-4\log_2\frac{1}{\eta}
-2\log_2 \frac{2}{\zeta'^2}-1\right],
\end{equation}
where 
$\zeta'=\left(\sqrt{\epsilon}-\eta+\frac{8f(p_\alpha, n)}{\sqrt{p_{\text{pass}}}}\right)/\left(3+\frac{5}{4p_{\text{pass}}}\right)$.

\end{thm}

\begin{proof}
The proof closely follows the one in Ref.~\cite{pfister2018}. 
Denote $\{Q_x\}_{x\in\chi}$ as the POVM measurement corresponding to homodyne detection in position basis and the measurement outcome is discretized 
in the set of alphabets $\chi$. Similarly, denote $\{P_x\}_{x\in\chi}$ as the POVM measurement corresponding to homodyne detection in momentum basis and measurement outcome is discretized in $\chi$.
For any random $\bm{z}\in\{0,1\}^{\otimes n}$, we define an isometry $V_{\bm{z}}:\mathcal H_{A^n}\rightarrow H_{A^n}\otimes H_{X^n}\otimes H_{X'^n}$ as an extension of the projective measurements on system $A^n$, where $X'^n$ are classical registers copying the information in $X^n$,
\begin{equation}
    V_{\bm{z}}: \ket{\psi}_{A^n}\rightarrow\sum_{\bm{x}\in \chi^{\otimes n}}  \Lambda_{\bm{z},\bm{x}}\ket{\psi}_{A^n}\ket{\bm{x}}_{X^n}\ket{\bm{x}}_{X'^n}
\end{equation}
where $\Lambda_{\bm{z},\bm{x}}=\otimes_{i=1}^n \Lambda_{z_i, x_i}$ and $\Lambda_{z,x}=\begin{cases}
Q_x & \text{if } z=0, \\
P_x & \text{if }  z=1.
\end{cases}$

As $\omega_{A^n X^n X'^n B^n E}$ can be obtained by applying an isometry on $\rho_{A^n B^n E}$, we have
\begin{equation}
    H_{\max}^{3\zeta+\zeta'+4\zeta''}(A^n|B^n)_\rho=H_{\max}^{3\zeta+\zeta'+4\zeta''}(A^n X^n X'^n|B^n)_\omega.
\end{equation}
Using Lemma~\ref{lemma:chainrule}, we get
\begin{equation}
    H_{\max}^{\zeta+\zeta'+2(\zeta+2\zeta'')}(A^n X^n X'^n|B^n)_\omega\le H_{\max}^{\zeta'}(X^n|A^n X'^nB^n)_\omega+H_{\max}^{\zeta+2\zeta''}(A^n X'^n|B^n)_\omega+\log\frac{2}{\zeta^2}.
\end{equation}
From the duality of min- and max-entropy~(\ref{eq:dualitySmoothEntropy}), we have
\begin{equation}
    H_{\max}^{\zeta'}(X^n|A^n X'^n B^n)_\omega=-H_{\min}^{\zeta'}(X^n|E)_\omega.
\end{equation}
Using Lemma~\ref{lemma:chainrule} again, we have
\begin{equation}
    H_{\max}^{\zeta+2\zeta''}(A^n X'^n|B^n)_\omega\le H_{\max}(A^n|X'^n B^n)_\omega+H_{\max}^{\zeta''}(X'^n|B^n)_\omega+\log\frac{2}{\zeta^2}.
\end{equation}
As $X$ and $X'$ stores the same information
\begin{equation}
    H_{\max}^{\zeta''}(X'^n|B^n)_\omega=H_{\max}^{\zeta''}(X^n|B^n)_\omega.
\end{equation}

Combining all above, we have for any $\zeta>0$ and $\zeta', \zeta''\ge 0$,
\begin{equation}
    H_{\max}^{3\zeta+\zeta'+4\zeta''}(A^n|B^n)_\rho \le H_{\max}(A^n|X'^n B^n)_\omega+H_{\max}^{\zeta''}(X^n|B^n)_\omega-H_{\min}^{\zeta'}(X^n|E)_\omega+2\log_2\frac{2}{\zeta^2}.
\end{equation}
We use the entropic uncertainty relation in Lemma~\ref{lemma:CVuncertainty} to obtain
\begin{equation}\label{appeq:quantumMaxEntropyToClassical}
    -H_{\max}^{3\zeta+\zeta'+4\zeta''}(A^n|B^n)_{\rho}\ge -n\log_2 c(d)-H_{\max}^{\zeta''}(X^n|B^n)_\omega-H_{\max}^{\zeta'-2\frac{f(p_{\alpha},n)}{\sqrt{p_{\operatorname{pass}}}}}(X^n|B^n)_\omega-2\log_2 \frac{2}{\zeta^2}.
\end{equation}
By setting $\zeta'=\frac{\zeta}{4 p_{\operatorname{pass}}}$ 
and $\zeta''=\zeta'-2\frac{f(p_{\alpha}, n)}{\sqrt{p_{\text{pass}}}}$,
using Lemma~\ref{lemma:boundMaxEntropy}, we have
\begin{equation}
    H_{\max}^{\zeta''}(X^n|B^n)_\omega=H_{\max}^{\zeta'-2\frac{f(p_{\alpha}, n)}{\sqrt{p_{\operatorname{pass}}}}}(X^n|B^n)_\omega\le n \log_2\gamma(t+\mu_0(\zeta)).
\end{equation}
By setting the relation
\begin{equation}
    3\zeta+\zeta'+4\zeta''=\sqrt{\epsilon/2}-\eta,
\end{equation}
we obtain
\begin{equation}
    \zeta=\left(\sqrt{\epsilon/2}-\eta+\frac{8f(p_\alpha, n)}{\sqrt{p_{\text{pass}}}}\right)/\left(3+\frac{5}{4p_{\text{pass}}}\right).
\end{equation}
When $\frac{\zeta}{4}-2f(p_{\alpha}, n)>0$, i.e.,
\begin{equation}
    0<\eta<\sqrt{\epsilon/2}-8f(p_{\alpha}, n)\left(3+\frac{5}{4p_{\text{pass}}}-\frac{1}{\sqrt{p_{\text{pass}}}}\right),
\end{equation} 
combining Lemma~\ref{lemma:capaitytoEntropy} and Eq.~(\ref{appeq:quantumMaxEntropyToClassical}), we get
\begin{equation}
    Q^{\epsilon}\gtrsim \sup_{\eta\in\left(0, \sqrt{\epsilon/2}-8f(p_{\alpha}, n)\left(3+\frac{5}{4p_{\text{pass}}}-\frac{1}{\sqrt{p_{\text{pass}}}}\right)\right)} \left[n\log_2 \frac{2\pi}{d^2}-2n \log_2\gamma(t+\mu_0(\zeta))
-2\log_2 \frac{2}{\zeta^2} +4\log_2 \eta-2\right].
\end{equation}
Using Lemma~\ref{lemma:distilledEntanglement}, we obtain a lower bound on the number of maximally entangled pairs which can be established by sending halves of two-mode squeezed vacuum states.

\end{proof}

\section{Proof of Theorem 2}
\label{sec:proofThm2}

We first present the protocol for independent and identical noises in the entanglement-based formalism instead of in the preparation-and-measurement  formalism as shown in the main text.
Alice prepares $n$ copies of two-mode squeezed vacuum states $\ket{\Psi_{\rho_{\text{th}(\bar{n})}}}$, feeds one party of each to a channel, and keeps the other party as reference modes.
For each copy, Alice and Bob choose a random phase shift operation $U\in\mathbb{U}(1)$, and apply operation $U^\dagger \otimes U$ at the reference mode and output mode.
After this symmetrization procedure, Alice and Bob both apply heterodyne measurements at the $n$ pairs of modes . 
Their measurement outcomes are denoted by $\bm{x}\in\mathbb{C}^{n}$ and $\bm{y}\in\mathbb{C}^{n}$, respectively. 
Later, we show that if the i.i.d noisy channel commutes with any phase rotation operation, then this symmetrization procedure is unnecessary to perform. 

Based on the measurement outcomes $\bm{x}$ and $\bm{y}$ as well as error probability $\delta$, Alice and Bob calculate 
\begin{align*}
   &\sigma_{\max}:=\frac{ ||\bm{y}||^2}{2(k-\sqrt{2k\ln1/\delta})}-1/2, \\
  &\gamma_{\min}:= \frac{||\bm{x}||^2+||\bm{y}||^2+2\bm{x}^\top\bm{y}}{4(k+\sqrt{2k\ln 2/\delta}+\ln2/\delta)}-\bar{n}-
    \frac{ ||\bm{y}||^2}{4(k-\sqrt{2k\ln1/\delta})}-3/4.
\end{align*} 
If the parameters satisfy $\sigma_{\max}\le a$ and $\gamma_{\min}\ge c$, then Alice and Bob pass the test. Otherwise, they abort the protocol.

Before we prove Theorem~2, we present a useful lemma.

\begin{lemma}[Asymptotic equipartition property for post-selected CV states~\cite{furrer2011}]\label{lemma:AEPCV}
Let $\sigma\in \mathcal H_A\otimes\mathcal H_B$ such that the von Neumann entropy $H(A)_{\sigma}$ is finite. For any $\epsilon>0$ and $n>\frac{8}{5}\log\frac{2}{\epsilon^2}$, we have 
\begin{equation*}
H_{\max}^{\epsilon}(A^n|B^n)_{\sigma^{\otimes n}} \le n H(A|B)_{\sigma}+4\sqrt{n}\log\nu \sqrt{\log\frac{2}{\epsilon^2}}
\end{equation*}
where $H(A|B)_{\sigma}=H(AB)_{\sigma}-H(B)_{\sigma}$ and $\nu:=\sqrt{2^{-H_{\min}(A|B)_{\sigma}}}+\sqrt{2^{H_{\max}(A|B)_{\sigma}}}+1$.
\end{lemma}

For i.i.d noisy channels, we suppose $\mathcal E=\Lambda^{\otimes n}$ and $\sigma=\Lambda\otimes \mathds{1} (\ket{\Psi_{\rho_{\text{th}(\bar{n})}}}\bra{\Psi_{\rho_{\text{th}(\bar{n})}}})$.
Using the fact that $H_{\min}(A|B)_{\sigma}\ge -2\log\tr(\sqrt{\sigma_A})$, $H_{\max}(A|B)_{\sigma}\le 2\log\tr(\sqrt{\sigma_A})$, and $\sigma_A=\rho_{\text{th}}(\bar{n})$,
we have
$\nu \le 2\sqrt{2^{2\log(\tr\sqrt{\sigma_A})}}+1= 2\sqrt{2^{2\log(\sqrt{1+\bar{n}}+\sqrt{\bar{n}})}}+1= 2\sqrt{1+\bar{n}}+2\sqrt{\bar{n}}+1$.
Using Lemma~\ref{lemma:AEPCV}, we have
\begin{align*}
H_{\max}^{\epsilon}(A^n|B^n)_{\sigma^{\otimes n}} \le&  n H(A|B)_{\sigma}+4\sqrt{n}\log(2\sqrt{1+\bar{n}}+2\sqrt{\bar{n}}+1) \sqrt{\log\frac{2}{\epsilon^2}}.
\end{align*}

After the symmetrization procedure $\sigma\rightarrow \tilde{\sigma}:=\int_{U\in\mathbb{U}(1)}dU U^\dagger\otimes U\sigma U\otimes U^\dagger$, the covariance matrix of $\tilde{\sigma}$ is
\begin{equation*}
    \begin{pmatrix}
      2\bar{n}+1 & 0 & \Sigma_c & \Sigma_d \\
      0 & 2\bar{n}+1 & \Sigma_d & -\Sigma_c\\
      \Sigma_c & \Sigma_d & \Sigma_b & 0 \\
      \Sigma_d  & -\Sigma_c & 0 & \Sigma_b
    \end{pmatrix}.
\end{equation*}
We find that the symplectic eigenvalues of the above matrix only depend on $\bar{n}$, $\Sigma_b$ and $\Sigma_c^2+\Sigma_d^2$.
 Fixing $\bar{n}$, $\Sigma_b$ and $\Sigma_c$, $H(A|B)_{\tilde{\sigma}}$ is maximized by minimizing $\Sigma_c^2+\Sigma_d^2$, which is achieved when $\Sigma_d=0$. 
It is easy to find that $H(A|B)_{\tilde{\sigma}}$ keeps increasing, when we raise $\Sigma_b$ and reduce $\Sigma_c$, because the noise within system $B$ is increased while the correlation between system $A$ and system $B$ decreases. Thus, an upper bound of $\Sigma_b$, together with a lower bound of $\Sigma_c$, yields an upper bound on $H(A|B)_{\tilde{\sigma}}$.

Suppose the channel $\Lambda(\cdot)$ commutes with any phase rotation operation $U\cdot U^\dagger$, then $\tilde{\sigma}=\mathds{1}\otimes \Lambda (\int_{U\in\mathbb{U}(1)}dU U^\dagger\otimes U \ket{\Psi_{\rho_{\text{th}(\bar{n})}}}\bra{\Psi_{\rho_{\text{th}(\bar{n})}}} U\otimes U^\dagger)$. Note that any phase rotation $U^\dagger \otimes U$ keeps a two-mode squeezed vacuum state $\ket{\Psi_{\rho_{\text{th}(\bar{n})}}}$ invariant. Thus, $\tilde{\sigma}=\sigma$, which implies that the symmetrization procedure does not need to be really performed.

Now we present how to obtain confidence intervals of variance $\Sigma_b$ and covariance $\Sigma_c$ from the finite measurement outcomes $\bm{x}$ and $\bm{y}$.
To achieve this goal, we consider the random variables of Alice and Bob' measurement outcomes as $\beta_A\in\mathbb C$ and $\beta_B\in\mathbb C$, respectively, which both follow Gaussian distributions. 
Then the covariance matrix of $\beta_A$ together with $\beta_B$ is
\begin{equation*}
    \begin{pmatrix}
      2\bar{n}+3/2 & 0 & \Sigma_c & \Sigma_d \\
      0 & 2\bar{n}+3/2 & \Sigma_d & -\Sigma_c\\
      \Sigma_c & \Sigma_d & \Sigma_b+1/2 & 0 \\
      \Sigma_d  & -\Sigma_c & 0 & \Sigma_b+1/2
    \end{pmatrix}.
\end{equation*}
This is because the fact that hetorodyne measurement combines the signal mode with a vacuum state by a balanced beam splitter and homodyne position and momentum of two resulting modes. 

From the definition of Chi-squared distribution, it is easy to see that $\frac{ ||\bm{y}||^2}{\Sigma_b+1/2}$ is a random variable following the Chi-squared distribution with $2k$ degrees. 
Then let us first introduce a concentration inequality for the Chi-squared distribution.
\begin{lemma}[Concentration inequality of Chi-squared distribution~\cite{laurent2000}]
Suppose variable $X$ follows the Chi-squared distribution with $n$ degrees. We have the following inequalities of probabilities, for any $x>0$,
\begin{align*}
    &\text{Pr}(X-n\ge 2\sqrt{nx}+ 2x)\le \text{e}^{-x}, \\
    &\text{Pr}(n-X\ge 2\sqrt{nx})\le \text{e}^{-x}.
\end{align*}
By setting $\delta=\text{e}^{-x}$, the above inequalities are transformed to
\begin{align}\label{eq:chi2inequal1}
    &\text{Pr}(X\ge n+2\sqrt{n\ln1/\delta}+2\ln1/\delta)\le \delta \\
    \label{eq:chi2inequal2}
    &\text{Pr}(X\le n-2\sqrt{n\ln1/\delta})\le \delta.
\end{align}
\end{lemma}

Using (\ref{eq:chi2inequal2}), we have
\begin{equation*}
    \text{Pr}\left(\frac{ ||\bm{y}||^2}{\Sigma_b+1/2}\le 2k-2\sqrt{2k\ln1/\delta}\right)\le \delta,
\end{equation*}
which is equivalent to
\begin{equation*}
    \text{Pr}\left(\Sigma_b \ge \frac{ ||\bm{y}||^2}{2(k-\sqrt{2k\ln1/\delta})}-1/2\right)\le \delta.
\end{equation*}
This is to say, with error probability at most $\delta$, the true variance satisfies
\begin{equation}\label{eq:boundsigmab}
    \Sigma_b \le \frac{ ||\bm{y}||^2}{2(k-\sqrt{2k\ln1/\delta})}-1/2.
\end{equation}

On the other hand, we consider the combination $\bar{\beta_A}+\beta_B$. It can be seen that both the real and imaginary part of $\bar{\beta_A}+\beta_B$ follow a Gaussian distribution with variance $2\bar{n}+3/2+\Sigma_b+1/2+2\Sigma_c=2\bar{n}+\Sigma_b+2\Sigma_c+2$.
Hence, $\frac{||\bar{\bm{x}}+\bm{y}||^2}{2\bar{n}+\Sigma_b+2\Sigma_c+2}$ follows the Chi-squared distribution with $2k$ degrees.
Using (\ref{eq:chi2inequal1}), we obtain
\begin{equation*}
    \text{Pr}\left(\frac{||\bar{\bm{x}}+\bm{y}||^2}{2\bar{n}+\Sigma_b+2\Sigma_c+2}\ge 2k+2\sqrt{2k\ln 1/\delta}+2\ln 1/\delta\right)\le \delta.
\end{equation*}
Using the relation $||\bar{\bm{x}}+\bm{y}||^2=||\bm{x}||^2+||\bm{y}||^2+2\bm{x}^\top\bm{y}$, the above inequality can be transformed to
\begin{equation*}
    \text{Pr}\left(\Sigma_c \le \frac{||\bm{x}||^2+||\bm{y}||^2+2\bm{x}^\top\bm{y}}{4(k+\sqrt{2k\ln 1/\delta}+\ln1/\delta)}-\bar{n}-\Sigma_b/2-1\right)\le \delta
\end{equation*}
That is with error probability at most $\delta$, the covariance is lower bounded by
\begin{equation*}
    \Sigma_c \ge \frac{||\bm{x}||^2+||\bm{y}||^2+2\bm{x}^\top\bm{y}}{4(k+\sqrt{2k\ln 1/\delta}+\ln1/\delta)}-\bar{n}-\Sigma_b/2-1.
\end{equation*}
Combining the fact in (\ref{eq:boundsigmab}), and the union bound, we obtain both the upper bound of variance $\Sigma_b$ the lower bound of covariance $\Sigma_c$ with error probability at most $\delta$,
\begin{align*}
&\Sigma_b \le \frac{ ||\bm{y}||^2}{2(k-\sqrt{2k\ln2/\delta})}-1/2, \\
 &   \Sigma_c \ge \frac{||\bm{x}||^2+||\bm{y}||^2+2\bm{x}^\top\bm{y}}{4(k+\sqrt{2k\ln 2/\delta}+\ln2/\delta)}-\bar{n}-
    \frac{ ||\bm{y}||^2}{4(k-\sqrt{2k\ln2/\delta})}-3/4.
\end{align*}

\begin{thm}
If the conditions $\sigma_{\max}\le a$ and $\gamma_{\min}\ge c$ are satisfied, then with
error probability less than $\delta$,
 the one-shot quantum capacity corresponding to each mode of the $k$ channels uses is bounded by
\begin{align*}
    \frac{Q^{\epsilon}}{k}\ge &\max\left\{0, g(a)-g(\nu_1)-g(\nu_2)+\frac{1}{k}\sup_{\eta\in\left (0, \sqrt{\epsilon/2}\right)} 
    \left[-4\sqrt{k}\log(2\sqrt{1+\bar{n}}+2\sqrt{\bar{n}}+1) \sqrt{\log\frac{2}{(\sqrt{\epsilon/2}-\eta)^2}}+4\log_2\eta-2 \right]\right\}.
\end{align*}
\end{thm}

\begin{proof}
 Using Lemma~\ref{lemma:capaitytoEntropy}, we have
\begin{equation}
   Q^{\epsilon}
   \ge \sup_{\eta\in \left(0, \sqrt{\epsilon/2}\right)} \left(-H_{\max}^{\sqrt{\epsilon/2}-\eta}(A^k|B^k)_{\sigma^{\otimes k}}+4\log_2\eta-2\right).
\end{equation}
Using Lemma~\ref{lemma:AEPCV}, we have
\begin{align*}
    Q^{\epsilon}\ge & \sup_{\eta\in \left(0, \sqrt{\epsilon/2}\right)} \left( -k H(A|B)_{\sigma}-4\sqrt{k}\log(2\sqrt{1+\bar{n}}+2\sqrt{\bar{n}}+1) \sqrt{\log\frac{2}{(\sqrt{\epsilon/2}-\eta)^2}}+4\log_2\eta-2\right).
\end{align*}
If the conditions $\sigma_{\max}\le a$ and $\gamma_{\min}\ge c$, then with probability at least $1-\delta$, $H(A|B)_\sigma$ is upper bounded by the conditional entropy of a Gaussian state with covariance matrix $\begin{pmatrix}
  (2\bar{n}+1)\mathds{1} & c\sigma_z \\
  c\sigma_z & a \mathds{1}
\end{pmatrix}$. That is
\begin{equation}
     H(A|B)_{\sigma}\le g(\nu_1)+g(\nu_2)-g(a),
\end{equation}
where $g(x):=\frac{x+1}{2}\log_2\frac{x+1}{2}-\frac{x-1}{2}\log_2\frac{x-1}{2}$, $\nu_1$ and $\nu_2$ are the symplectic eigenvalues of covariance matrix $\begin{pmatrix}
  (2\bar{n}+1)\mathds{1} & c\sigma_z \\
  c\sigma_z & a \mathds{1}
\end{pmatrix}$.
\end{proof}

\section{Estimating lower bounds on quantum capacity of qubit channels}

The protocol to estimate lower bounds on quantum capacities for i.i.d qubit channels is first preparing a maximally entangled state $\ket{\Psi_+}=\frac{1}{\sqrt{2}}(\ket{00}+\ket{11})$. 
Then Alice applies a quantum channel at one party of $\ket{\Psi_+}\bra{\Psi_+}$ and keeps the other party as a reference qubit. At output side, Bob randomly chooses to measure Pauli observable $\sigma_{B,i}\otimes\sigma_{A,j}$, where $i,j=0,1,2,3$ and $\sigma_{0,1,2,3}=\mathds{1}, \sigma_x, \sigma_y, \sigma_z$.
After $n$ rounds of measurements, following the theorem below, Alice and Bob can calculate a lower bound on quantum capacity.

\begin{lemma}[Fully quantum AEP~\cite{tomamichel2009}]\label{lemma:AEP}
For any $\sigma_{AB}$,
 \begin{equation}
     H_{\max}^\epsilon(A^n|B^n)_{\sigma^{\otimes n}}\le  n H(A|B)_{\sigma}+4\sqrt{n} \log_2 \mu \sqrt{\log_2 \frac{2}{\epsilon^2}}
 \end{equation}
 where $\mu \le \sqrt{2^{H_{\min}(A|B)_\sigma}}+\sqrt{2^{-H_{max}(A|B)_\sigma}}+1\le 2^{ d_A/2 +2}$.
\end{lemma}

\begin{lemma}[Confidence polytope of quantum tomography~\cite{wang2019}]
For $k$th ($0\le k\le d^4-1$) Pauli observable, denote the corresponding POVM by $\mathcal M_k:=\{ E_k^{(l)}\}_{l=0}^{d-1}$ on $\mathcal H_A\otimes \mathcal H_B$, where $l$ denotes the measurement outcome. After the measurements $\otimes_{k=0}^{d^2-1}\mathcal M_k^{\otimes n_k}$, for each $k$, the number of rounds of measurements getting outcome~$l$ is $n_k^l$.  The confidence interval of the state $\sigma\in \mathcal S(\mathcal H_A\otimes \mathcal H_B)$, with confidence level $1-\delta$, where $\delta=\sum_{k=0}^{d^2-1}\sum_{l=0}^{d-1} \delta_k^l$, is $\Gamma=\cap_{0\le k\le d^2-1, 0\le l\le d-1} \Gamma_{kl}$, where
\begin{equation}\label{eq:halfspace}
\Gamma_{kl}:=\left\{\rho\in \mathcal S(\mathcal H_A\otimes \mathcal H_B): \frac{n_k}{n}\tr\left(\rho E_k^{(l)}\right)\le \frac{n_k^l}{n}+\epsilon\left(n_k^l,\delta_k^l\right)\right\},
\end{equation}
Here $\epsilon\left(n_k^l,\delta_k^l\right)$ is the positive root of the equation
\begin{equation}
    D\left(\frac{n_k^l}{n}||\frac{n_k^l}{n}+\epsilon\right)=-\frac{1}{n}\log\delta_k^l,
\end{equation}
where $D(x||y)=x\log\frac{x}{y}+(1-x)\log\frac{1-x}{1-y}$.
\end{lemma}

\begin{thm} Suppose by applying quantum state tomography described above, we get a confidence region $\Gamma$. Then we have
\begin{equation}
   \frac{Q^{\epsilon}(\mathcal E)}{n}\ge  - \max_{\sigma_{AB}\in \Gamma} H(A|B)_{\sigma}+ \sup_{\eta\in \left(0,\sqrt{\epsilon/2}\right)} \frac{4}{n}\left[- (d_A/2 +2)\sqrt{n} \sqrt{\log_2 \frac{2}{(\sqrt{\epsilon/2}-\eta)^2}} +\log_2 \eta\right]-\frac{2}{n}.
\end{equation}
\end{thm}

One of our motivations to propose this protocol to estimate lower bounds on one-shot quantum capacities for i.i.d noisy channels is that the previous lower bound obtained by the protocol in Ref.~\cite{pfister2018} can be far from the optimal lower bound for some practically important i.i.d noisy channels. 
Particularly consider the following parametrized quantum channel
\begin{equation}\label{eq:parametrizedChannel}
    \mathcal E(\rho)=\sum_{i=1}^2 A_i\rho A_i^\dagger,
\end{equation}
where $A_1=\cos\alpha\ket{0}\bra{0}+\cos\beta\ket{1}\bra{1}$ and $A_2=\sin\beta\ket{0}\bra{1}+\sin\alpha\ket{1}\bra{0}$. When $\alpha=\beta$, the quantum channel is a dephasing channel and when $\beta=0$, the channel becomes a amplitude damping channel. Its quantum capacity is nonzero only when $\cos(2\alpha)/\cos(2\beta)>0$.

The detectable lower bound in our protocol asymptotically approaches coherent information
\begin{equation}\label{eq:ampDampingLowerBound}
    -H(A|B)_{\sigma}=h((\cos^2\alpha+\sin^2\beta)/2)+h((\sin^2\alpha+\sin^2\beta)/2).
\end{equation}
Fig.~\ref{fig:amplitudeDamping} shows the difference between the lower bound~(\ref{eq:ampDampingLowerBound}) and the one obtained using the method in Ref.~\cite{pfister2018}.  
As it shows, for i.i.d dephasing channels, our protocol, by estimating coherent information, provides the same lower bound on quantum capacity in the asymptotic limit. However, for i.i.d amplitude damping channels, our protocol outperforms the one in Ref.~\cite{pfister2018} asymptotically, providing a tighter lower bound on quantum capacities.
\begin{figure}
\begin{center}
\includegraphics[width=0.4\textwidth]{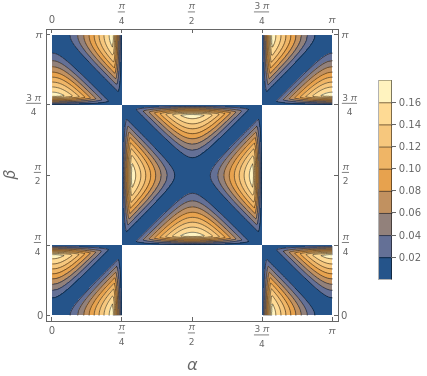}
\caption{The difference between the coherent information~(\ref{eq:ampDampingLowerBound}) and the detectable lower bound of quantum capacity in Ref.~\cite{pfister2018} for quantum channels in Eq.~(\ref{eq:parametrizedChannel}) within the region $\cos(2\alpha)/\cos(2\beta)>0$.}
\label{fig:amplitudeDamping}
\end{center}
\end{figure}

In the following, we extend the above result to general non-i.i.d scenario by using quantum de Finetti theorem.
We suppose $\rho_{A^{n+k} B^{n+k}}$ is an arbitrary state jointly at $A$ and $B$ with $n+k$ pairs of qubits/qudits. 
 As $\rho_{A^{n+k} B^{n+k}}$ is permutation-invariant, there always exists a purification $\rho_{A^{n+k}B^{n+k} E^{n+k}}\in \mathcal S(\text{Sym}\left((\mathcal H_A\otimes \mathcal H_B\otimes \mathcal H_E))^{\otimes n+k}\right)$, where $E\cong A\otimes B$.
\begin{lemma}[Exponential quantum de Finetti theorem~\cite{renner2008}]\label{lemma:deFinetti}
The trace distance between $\rho_{A^n B^n E^n}:=\tr_{A^k B^k E^k} \rho_{A^{n+k} B^{n+k} E^{n+k}}$ and a mixture of almost iid pure states $\tilde{\rho}^{\theta}\in \mathcal S(\text{Sym}(\mathcal H_{ABE}^{\otimes n},\ket{\theta}^{\otimes n-r}))$ can be bounded by 
\begin{equation}\label{eq:deFinetti}
 ||\rho_{ABE}^n -\int d\nu(\theta) \tilde{\rho}^{\theta}||_1\le 2 k^{d/2}\cdot \text{e}^{-\frac{k(r+1)}{2(n+k)}}
\end{equation}
where $\nu$ is a probability measure on $\mathcal H_{ABE}$ and $d=\text{dim}(\mathcal H_{ABE})$.
\end{lemma}
 For qubits, $d=2^4=16$ and the right hand side of Eq.~(\ref{eq:deFinetti}) becomes $2 k^8\cdot \text{e}^{-\frac{k(r+1)}{2(n+k)}}$.
 
The quantum asymptotic equipartition property~\cite{tomamichel2009}, shown in Lemma~\ref{lemma:AEP}, can be generalized to almost iid states as follows.
\begin{lemma}[fully quantum AEP for almost iid states]\label{lemma:AEPalmostiid}
Given $\tilde{\rho}^\theta:=\ket{\Psi_\theta}\bra{\Psi_\theta}$ has an almost iid structure, i.e., $\ket{\Psi_\theta}_{ABE}\in \mathcal Sym(\mathcal H_{ABE}^{\otimes n}, \ket{\theta}^{\otimes n-r})$, from the asymptotic equipartition property, we have
\begin{equation}
 -H^{\epsilon}_{\max}(A^n|B^n)_{\tilde{\rho}^\theta} \ge -(n-r) H(A|B)_{\ket{\theta}\bra{\theta}} -4\sqrt{n-r}\log \mu \sqrt{\log \frac{2}{\tilde{\epsilon}^2}}-n\cdot h(r/n)-r \log_2 d_A,
\end{equation}
where  $\tilde{\epsilon}\ge \frac{\epsilon^2}{6\cdot 2^{n \cdot h(r/n)}}$, 
$\mu \le \sqrt{2^{-H_{\min}(A|E)_{\ket{\theta}\bra{\theta}}}}+\sqrt{2^{H_{max}(A|E)_{\ket{\theta}\bra{\theta}}}}+1\le 2^{d_A/2 +1}+1$, where $d_A=\operatorname{dim}(\mathcal H_A)$. The above bound can be further simplified to
\begin{align}\notag
  -H^{\epsilon}_{\max}(A^n|B^n)_{\tilde{\rho}^\theta}  \ge&  (n-r) (H(B)_{\ket{\theta}\bra{\theta}}-H(AB)_{\ket{\theta}\bra{\theta}})\\
 &-4\sqrt{n-r}\log \mu \sqrt{2n h(r/n)-4\log \epsilon+2\log 6+1}-n h(r/n)-r \log_2 d_A.
\end{align}
\end{lemma}
The proof of this Lemma closely follows the idea in the proof of Theorem 4.4.1. in Ref.~\cite{renner2008}.
\begin{proof}
There exists a family of mutually orthonormal states $\{\ket{\psi_s}\}_{s\in S}$ on $Sym(\mathcal H_{ABE}^{\otimes n}, \ket{\theta}^{\otimes n-r})$ with $|S|\le 2^{nh(r/n)}$ such that $\ket{\Psi_\theta}=\sum_{s\in S}\gamma_s \ket{\psi_s}$ with $\sum_{s\in S}|\gamma_s|^2=1$.  Then the reduced state 
$\rho_{A^nE^n}=\tr_{B^n} \left(\ket{\Psi_\theta}\bra{\Psi_\theta}\right)$ and $\tilde{\rho}_{A^nE^n}^s=\tr_{B^n} \left(\ket{\psi_s}\bra{\psi_s}\right)$.
Another state is defined 
$\tilde{\rho}_{A^n E^n S}:=\sum_{s\in S} |\gamma_s|^2 \tilde{\rho}_{AE}^s \otimes \ket{s}\bra{s}$.
Then it has been shown that
\begin{align*}
H_{\min}^\epsilon(A^n|E^n)_{\rho}\ge H_{\min}^{\tilde{\epsilon}}(A^n|E^n S)_{\tilde{\rho}}-H_{max}(\tilde{\rho}_S)\\
\ge \min_{s\in S} H_{\min}^{\tilde{\epsilon}} (A^n| E^n)_{\tilde{\rho}^s}-n h(r/n),
\end{align*}
where $\tilde{\epsilon}=\frac{\epsilon^2}{6|S|}$, and we have used the fact that $H_{max}(\tilde{\rho}_S)=\log_2\text{rank}(\tilde{\rho}_S)=n h(r/n)$.

Without loss of generality, $\ket{\psi_s}=\ket{\theta}^{n-r}\otimes \ket{\hat{\psi}_s}$ for some $\ket{\hat{\psi}_s}\in \mathcal H_{ABE}^{\otimes r}$. Then
\begin{align*}
\tilde{\rho}_{A^n E^n}^s&=\tr_{B^n} \left( \ket{\theta}\bra{\theta}^{\otimes n-r}\otimes \ket{\hat{\psi}_s}\bra{\hat{\psi}_s} \right)\\
&=(\tr_{B}\ket{\theta}\bra{\theta})^{\otimes n-r}\otimes \tr_{B^{r}}\ket{\hat{\psi}_s}\bra{\hat{\psi}_s}
\end{align*}
Denote $\hat{\rho}_{A^r E^r}^s=\tr_{B^{r}}\ket{\hat{\psi}_s}\bra{\hat{\psi}_s}$ and $\sigma_{AE}=\tr_B \ket{\theta}\bra{\theta}$. By superadditivity of min-entropy, we have
\begin{equation*}
H_{\min}^{\tilde{\epsilon}}( A^n|E^n)_{{\tilde{\rho}}^s}\ge  H_{\min}^{\tilde{\epsilon}}(A^{n-r}| E^{n-r})_{\sigma^{\otimes n-r}}+H_{\min} (A^r|E^r)_{\hat{\rho}^s}.
\end{equation*}
Using the asymptotic equipartition property for iid states~\cite{tomamichel2009} that is $H_{\min}^\epsilon(A^{n-r}|E^{n-r})_{\sigma^{\otimes n-r}}\ge (n-r) H(A|E)_{\sigma}-\sqrt{n-r}\delta(\epsilon, \mu)$,
where $\delta(\epsilon, \mu)=4\log_2 \mu \sqrt{\log_2 \frac{2}{\epsilon^2}}$, and $H_{\min} (A^r|E^r)_{\hat{\rho}^s} \ge  -2\log_2 \tr\sqrt{\hat{\rho}_{A^r}^s}\ge -r\log_2 d_A$, we obtain for any $s$,
\begin{equation*}
H_{\min}^{\tilde{\epsilon}}(A^n | E^n)_{{\tilde{\rho}}^s}\ge (n-r) H(A|E)_{\sigma}-\sqrt{n-r}\delta(\tilde{\epsilon}, \mu)-r\log_2 d_A.
\end{equation*}
Hence, we have
\begin{equation*}
H_{\min}^\epsilon(A| E)_{\rho_{A E}}\ge   (n-r) H(A|E)_{\sigma_{AE}}-\sqrt{n-r}\delta(\tilde{\epsilon}, \mu)-r\log_2 d_A -n h(r/n).
\end{equation*}
From duality of smooth min- and max-entropy, we obtain the result.
\end{proof}

\begin{lemma}[polytope confidence interval for almost iid state quantum tomography]\label{lemma:polytope}
$\ket{\Psi_\theta}\in \mathcal Sym(\mathcal H_{ABE}^{\otimes n}, \ket{\theta}^{\otimes n-r})$, where $r<n/2$.
Suppose we apply local Pauli measurements at input $A$ and output $B$.
For $k$th ($0\le k\le d^2-1$) Pauli observable, denote the corresponding POVM by $\mathcal M_k:=\{ E_k^{(l)}\}_{l=0}^{d-1}$ on $\mathcal H_A\otimes \mathcal H_B$, where $l$ denotes the measurement outcome. After the measurements $\otimes_{k=0}^{d^2-1}\mathcal M_k^{\otimes n_k}$, for each $k$, the number of rounds of measurements getting outcome~$l$ is $n_k^l$.  The confidence interval of state $\rho_{AB}=\tr_E\ket{\theta}\bra{\theta}$, with confidence level $1-\delta$, where $\delta=\sum_{k=0}^{d^2-1}\sum_{l=0}^{d-1} \delta_k^l$, is $\Gamma=\cap_{0\le k\le d^2-1, 0\le l\le d-1} \Gamma_{kl}$, where
\begin{equation}
\Gamma_{kl}:=\left\{\rho\in \mathcal S(\mathcal H_{AB}): \tr\left(\rho E_k^{(l)}\right)\le \frac{n_k^l}{n_k}+\frac{n}{n_k}\sqrt{\frac{\log_2 1/\delta_k^l}{n}+h(r/n)+\frac{2}{n} \log_2(n/2+1) }\right\}.
\end{equation}
\end{lemma}

\begin{proof}
The proof combines the idea of confidence polytope in quantum tomography~\cite{wang2019} with the statistical properties of almost iid states~\cite{renner2008}. The POVM measurements at $\mathcal H_{AB}$ can be easily extended to $\mathcal H_{ABE}$ by denoting $\tilde{\mathcal M_k}:=\{\tilde{E}_k^{(l)}\}_{l=0}^{d-1}$, where $\tilde{E}_k^{(l)}:=E_k^{(l)}\otimes \mathds{1}_E$. 
A renormalized POVM on $\mathcal H_{ABE}$ is $\tilde{\mathcal M}:=\{\frac{n_k}{n} \tilde{E}_k^{(l)}\}_{k=0,l=0}^{d^2-1,d-1}$.

Then we consider POVM $\left\{\frac{n_k}{n} \tilde{E}_k^{(l)}, \mathds{1}_{ABE}-\frac{n_k}{n} \tilde{E}_k^{(l)}\right\}$. 
Using Theorem 4.5.2 in Ref.~\cite{renner2008}, we obtain for each $k$ and $l$,
\begin{equation}
    \text{Pr}\left(\left|\braket{\theta|\tilde{E}_k^{(l)} |\theta} -\frac{n_k^l}{n_k}\right|>\frac{n}{n_k}\sqrt{ \frac{\log_2(1/\delta_k^l)}{n_k} +h(r/n)+\frac{2}{n}\log_2(n_k/2+1)}\right)\le \delta_k^l.
\end{equation}
By noting that $\tr\left((\tr_{E}\ket{\theta}\bra{\theta}) E_k^{(l)}\right)=\braket{\theta|\tilde{E}_k^{(l)}|\theta}$, we get
\begin{equation}
    \text{Pr}\left( \tr\left(\rho E_k^{(l)}\right)> \frac{n_k^l}{n_k}+\frac{n}{n_k}\sqrt{\frac{\log_2 1/\delta_k^l}{n}+h(r/n)+\frac{2}{n} \log_2(n/2+1) }\right)\le\delta_k^l.
\end{equation}
Finally, the union bound indicates that $\sigma\in \cap_{0\le k\le d^2-1, 0\le l\le d-1} \Gamma_{kl}$ with probability at least $1-\sum_{k=0}^{d^2-1}\sum_{l=0}^{d-1} \delta_k^l$.
\end{proof}

\begin{thm}
Given a quantum channel $\mathcal E^{n+k}: \mathcal H_{A'}^{\otimes n+k}\rightarrow \mathcal H_B^{\otimes n+k}$. We feed one party of the maximally entangled state at each input and keep the other party as a reference system. We randomly abandon $k$ outputs and 
denote the channel corresponding to the other $n$ inputs and $n$ outputs by $\mathcal E^n$. For any error $\epsilon/2>\epsilon':=2 k^{d/2} \text{e}^{-\frac{k(r+1)}{2(n+k)}}$, we have the lower bound of one-shot quantum capacity
of $\mathcal E^n$
\begin{align}
Q^{\epsilon}(\mathcal E^n)\notag
\ge & \max\Big\{0, \sup_{\eta\in \left(0,\sqrt{\epsilon/2}-\sqrt{\epsilon'}\right)}\Big[-4\sqrt{n-r}\log(2\sqrt{2}+1) \sqrt{2n h(r/n)-4\log (\sqrt{\epsilon/2}-\eta-\sqrt{\epsilon'}) +2\log 6+1}\\
&+4\log_2 \eta \Big]-n h(r/n)-r+ (n-r)\min_{\sigma\in \Gamma} (H(B)_\sigma-H(AB)_\sigma)  -2 \Big\}.
\end{align}
\end{thm}
\begin{proof}
Lemma~\ref{lemma:capaitytoEntropy} tells us that $Q^{\epsilon}(\mathcal E^n)$ can be bounded below by a function of smooth max-entropy $H_{\max}^{\sqrt{\epsilon/2}-\eta}(A^n|B^n)_{\rho}$ optimized over $\eta\in (0,\sqrt{\epsilon/2})$, where $\rho^n$ is the state at the $n$ output qubits and the associated~$n$ ancillary qubits. The smooth max-entropy itself is a minimum value within a neighborhood $\mathcal B^{\sqrt{\epsilon/2}-\eta}(\rho_{A^n B^n})$.
As Lemma~\ref{lemma:deFinetti}, together with the fact that partial trace can only reduce trace distance, implies that $\rho_{A^nB^n}$ is close to an unknown almost iid state $\tilde{\rho}_{A^n B^n}$,
we can use the minimum value over a smaller neighborhood around $\tilde{\rho}_{A^n B^n}$, which is a subset of $\mathcal B^{\sqrt{\epsilon/2}-\eta}(\rho_{A^n B^n})$, to obtain an upper bound on $H_{\max}^{\sqrt{\epsilon/2}-\eta}(A^n|B^n)_{\rho}$.

Using the triangle inequality of purified distance~\cite{PhysRevA.71.062310}, we have, for any $\rho'_{A^n B^n}\in \mathcal S(\mathcal H_{A^n B^n})$,
\begin{equation}
\mathcal P(\rho_{A^n B^n},\rho'_{A^n B^n} )\le  \mathcal P\left(\rho_{A^n B^n}, \int d\nu(\theta) \tilde{\rho}_{A^n B^n}^{\theta}\right) + \mathcal P\left(\rho'_{A^n B^n},\int d\nu(\theta) \tilde{\rho}_{A^n B^n}^{\theta}\right).
\end{equation}
To make sure $\mathcal P(\rho_{A^n B^n},\rho'_{A^n B^n})\le \sqrt{\epsilon/2}-\eta$, as $\mathcal P(\rho_{A^n B^n}, \int d\nu(\theta) \tilde{\rho}_{A^n B^n}^{\theta})\le \sqrt{\lambda}$ with $\epsilon':=2 k^{8} \text{e}^{-\frac{k(r+1)}{2(n+k)}}$,
we only need to set $\mathcal P(\rho'_{A^n B^n},\int d\nu(\theta) \tilde{\rho}_{A^n B^n}^{\theta})\le \sqrt{\epsilon/2}-\eta-\sqrt{\epsilon'}$. Hence 
using both Lemma~\ref{lemma:AEPalmostiid} and Lemma~\ref{lemma:polytope}, we get a lower bound, when $\eta<\sqrt{\epsilon/2}-\sqrt{\epsilon'}$,
\begin{align*}
&-H_{\max}^{\sqrt{\epsilon/2}-\eta}(A^n|B^n)_\rho\ge -H_{\max}^{\sqrt{\epsilon/2}-\eta-\sqrt{\epsilon'}} (A^n|B^n)_{\tilde{\rho}}\\
\ge &-4\sqrt{n-r}\log(2\sqrt{2}+1) \sqrt{2n h(r/n)-4\log (\sqrt{\epsilon/2}-\eta-\sqrt{\epsilon'}) +2\log 6+1}\\
&-n h(r/n)-r + (n-r) \min_{\sigma\in\Gamma}(H(B)_\sigma-H(AB)_\sigma) ,
\end{align*}
and hence using Lemma~\ref{lemma:capaitytoEntropy} we get the result.
\end{proof}

\end{widetext}

\end{document}